\documentclass[epj]{svjour}
\usepackage[utf8]{inputenc}
\usepackage{graphics}
\usepackage{graphicx}
\usepackage{physics}
\usepackage{multirow}
\usepackage{booktabs}
\usepackage{breqn}
\usepackage[switch,columnwise,displaymath,mathlines]{lineno}
\usepackage{amsmath}
\usepackage{amssymb}
\begin{document}
\hugehead

\title{Precision resonance energy scans with the PANDA experiment at FAIR}
\subtitle{Sensitivity study for width and line shape measurements of the X(3872)}
%
%
\author{
{\large The PANDA Collaboration}\\ \\
G.~Barucca\inst{1} \and 
F.~Davì\inst{1} \and 
G.~Lancioni\inst{1} \and 
P.~Mengucci\inst{1} \and 
L.~Montalto\inst{1} \and 
P. P.~Natali\inst{1} \and 
N.~Paone\inst{1} \and 
D.~Rinaldi\inst{1} \and 
L.~Scalise\inst{1} \and 
W.~Erni\inst{2}
% ...
}
\institute{
Università Politecnica delle Marche-Ancona, {Ancona}, Italy \and 
Universität Basel, {Basel}, Switzerland
% ...
}

%
\offprints{F.\;Nerling (f.nerling@gsi.de)}          
%
%

\date{Received: date / Revised version: date}
%
\abstract{
This paper summarises a comprehensive Monte Carlo simulation study for precision resonance energy scan measurements. Apart from the proof of principle for natural width and line shape measurements of very narrow resonances with PANDA, the achievable sensitivities are quantified for the concrete example of the charmonium-like $X(3872)$ state discussed to be exotic, and for a larger parameter space of various assumed signal cross-sections, input widths and luminosity combinations. PANDA is the only experiment that will be able to perform precision resonance energy scans of such narrow states with quantum numbers of spin and parities that differ from $J^{PC} = 1^{--}$.
\PACS{
      {01.52.+r}{International laboratory facilities} \and
      {13.25.-k}{Hadron decays, mesons}  \and
      {13.75.-n}{Hadrons, interactions induced by low and intermediate energy}  \and
      {14.40.Rt}{Exotic mesons}   \and
      {14.40.-n}{Hadrons, properties of mesons}   \and
      {14.40.Pq}{Quarkonia heavy quarkonia}   \and
      {25.40.Ny}{Resonance reactions, nucleon-induced} \and
      {25.43.+t}{Antiproton-induced reactions}  
     } 
} 
\authorrunning{K.~G\"otzen, R.~Kliemt, F.~Nerling and K.~Peters, et al.}
\titlerunning{Sensitivity study for resonance energy scans with PANDA} 
\maketitle
\section{Introduction}
\label{intro}
Since the beginning of the millennium, many charmonium-like states, the so-called $XYZ$ states, have been observed experimentally, showing characteristics different from the predictions of conventional charmonium states predicted by potential models, and being therefore largely 
discussed to be of exotic nature. The first and most intriguing one is the famous $X(3872)$ that was discovered
by the Belle Collaboration in $B^{\pm} \rightarrow K^{\pm}X(3872), X(3872)\rightarrow \pi^+\pi^-J/\psi$ in 2003~\cite{BelleX3872}. This state has subsequently been confirmed by other experiments~\cite{X3872others_2_CDF,X3872others_3_D0,X3872others_4_BaBar,X3872others_5_LHCb}. Further measurements indicate the di-pion system in the $J/\psi\,\pi^+\pi^-$ to originate from the $\rho^0(770)$~\cite{X3872_JpsiRho_CDF}, 
implying an unusually strong isospin violation for {\it e.g.} the interpretation as a conventional charmonium state. The vector states $Y(4260)$, $Y(4360)$ and $Y(4660)$ have been discovered by the BaBar, Belle and CLEO Collaborations in the decays to the final states $J/\psi\,\pi^+\pi^-$ and $\psi(3686)\,\pi^+\pi^-$, comprising low-mass charmonia~\cite{Y4260_1_BaBar,Y4360_1_Belle,Y4660_1_CLEO,Y4360_2_BaBar,Y4xxx_2_Belle}. 
The manifestly exotic charged char{-}monium-like states such as the $Z_{\rm c}(4430)$ have been observed by different experiments~\cite{Zc4430_1_Belle,Zc4430_2_LHCb}, 
and particularly for the $Z_{\rm c}(3900)$ and the $Z_{\rm c}(4020)$~\cite{Zc3900_1_BESIII,Zc3900_2_Belle,Zc4020_1_CLEO,Zc4020_2_BESIII,Zc3885_1_BESIII,Zc4025_1_BESIII}, also the neutral isospin partners have been found~\cite{Zc3900_Neutral_BESIII,Zc3885_Neutral_BESIII,Zc4020_Neutral_1_BESIII,Zc4025_Neutral_1_BESIII}, {\it cf.}~\cite{PDG18}. For a recent overview on these experimental findings and the resulting $XYZ$ puzzle, see {\it e.g.}~\cite{ReviewPapaerMitcheletAl,Olsen:2017bmm}.

Various interpretations on the nature of the $XYZ$ 
\newline
states have been proposed, including molecular, hybrid, multi-quark states and also other explanations, such as threshold enhancements and some other configurations, see {\it e.g.}~\cite{XYZinterpretations,Guo:2017jvc,Esposito:2016noz}. The nature of these states is, however, still unclear. Especially for the $X(3872)$, the to-date measured mass is indistinguishable from the $DD^*$ threshold~\cite{PDG15}. It is even not clear yet, whether it lays beneath or above this threshold, and due to the rather narrow natural decay width, merely an experimental upper limit of 1.2\,MeV at a 90\% confidence level exists~\cite{belle_Xwidth}. To understand the nature and distinguish between the various theoretical models, an absolute width measurement with sub-MeV resolution is required for this $J^{PC}=1^{++}$ state~\cite{LHCbXJPC}. For states with $J^{PC}$ different from $1^{--}$, such a precision measurement can only be performed with an antiproton-proton ($\bar{p}p$) experiment such as PANDA, {\it cf. e.g.}~\cite{PandaPhysBook09}.
\begin{figure*}[tp!]
\centering
  \includegraphics[clip,trim= 0 20 35 100, width=0.8\textwidth]{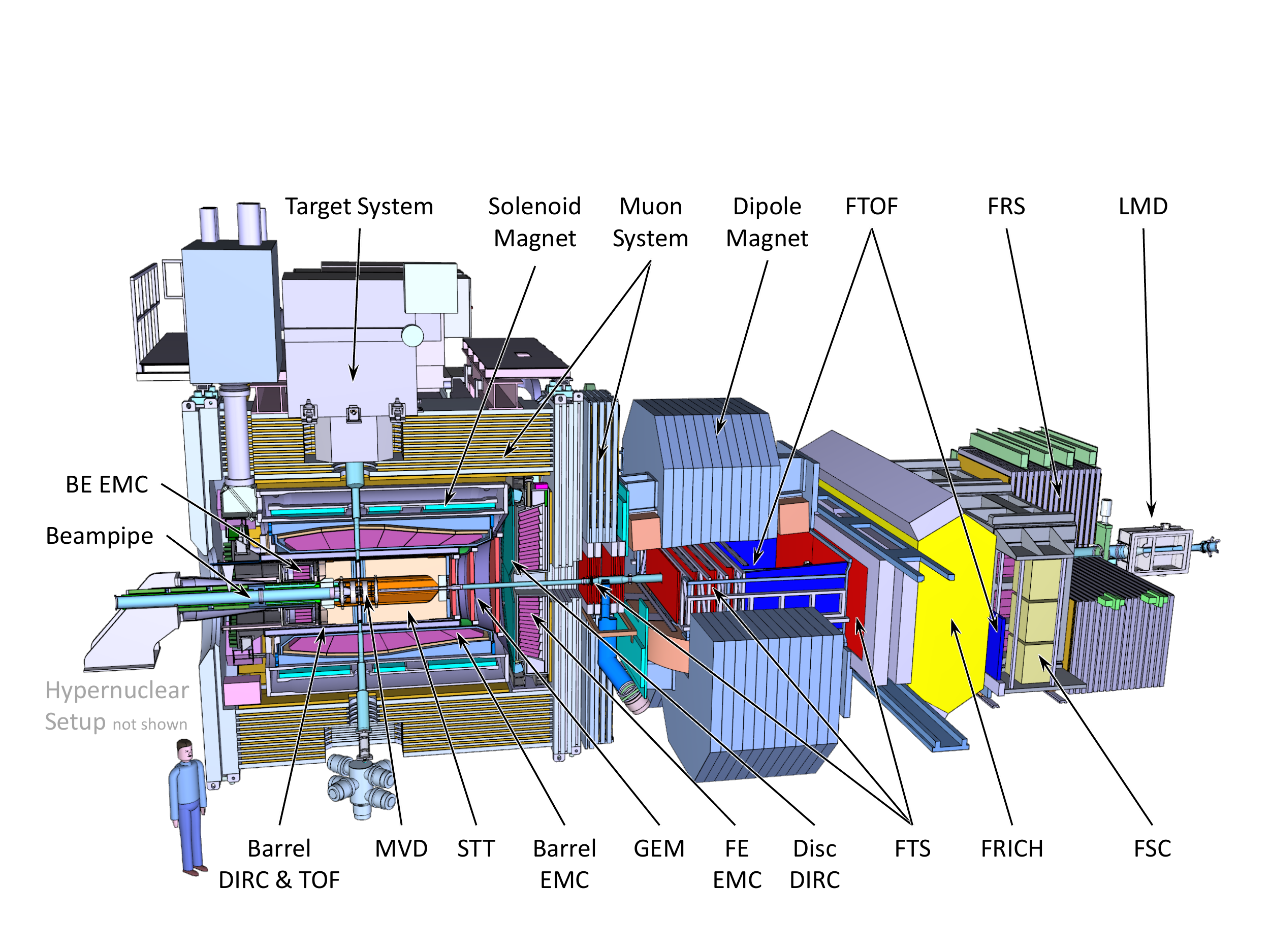}
  \caption{\label{fig:panda_setup} The proposed PANDA experimental setup.}
\end{figure*}

\section{The PANDA experiment at FAIR}
\label{sec:PandaFair}
The PANDA (antiProton ANnihilation in DArmstadt) experiment~\cite{PandaPhysBook09} will be located at the FAIR (Facility for Antiproton and Ion Research) complex under construction in Darmstadt, Germany. The physics programme is dedicated to hadron physics. Apart from hadron spectroscopy in the charmonium and light quark regime, nucleon structure and hypernuclear physics will be studied. Moreover, {\it e.g.} in-medium modifications of charm in nuclear matter and physics of strangeness production are part of the programme. 

\subsection{The Facility for Antiproton and Ion Research}
\label{subsec:FAIR}
The FAIR accelerator complex is built to extend the existing GSI (Helmholtzzentrum f\"ur Schwerionenforschung GmbH) facilities. It will provide particle beams for four main experimental pillars, one of which is the PANDA experiment dedicated to hadron physics. At FAIR, a new proton LINAC will pre-accelerate protons to 70\,MeV and feed them to the existing SIS18 synchrotron ring with a bending power of 18\,Tm that will further accelerate them to 3.7\,GeV/$c$ and inject them subsequently into a new, larger synchrotron SIS100 (bending power of 100\,Tm), further accelerating them to about 30\,GeV/$c$. 

The 30 GeV/$c$ proton beam will hit a copper target acting as the antiproton production target. Time averaged production rates in the range of $5.6 \times 10^6$ to $10^7$ antiprotons are expected. Magnetic horns are then used to filter the antiprotons of 3.7\,GeV/$c$, which are then collected and phase-space cooled in the Collector Ring (CR), then transferred and, in the final setup stored in the Recycled Experimental Storage Ring (RESR). In the initial start-up phase of FAIR operation, without the RESR, the accumulation of the antiprotons will be done in the High Energy Storage Ring (HESR), resulting in a reduced luminosity being about a factor of ten lower than the nominal design value. 

Finally, the 3.7\,GeV/$c$ antiprotons are injected in the HESR, equipped with stochastic cooling, where they are collected and the beam is cooled. Here, they can be de-accelerated or further accelerated, covering a range of deliverable antiproton momenta for the PANDA fixed-target experiment between 1.5\,GeV/$c$ and 15\,GeV/$c$. The full setup is designed to provide an antiproton beam of up to $10^{11}$ antiprotons per filling and an instantaneous peak luminosity reaching up to $2\times 10^{32}$ cm$^{-2}$s$^{-1}$. This allows for the accumulation of an integrated luminosity of 2\,fb$^{-1}$ in about five months.
 
\subsection{The PANDA detector}
\label{subsec:Detector}
The proposed PANDA multi-purpose detector~\cite{PandaPhysBook09,TDRs} as shown in Fig.\,\ref{fig:panda_setup} will be located at the HESR. It consists of the target spectrometer surrounding the target area and the forward spectrometer for the detection of particles produced in the forward direction and thus being detected at central rapidities. Almost $4\pi$ geometrical acceptance will be covered, as particularly needed for measurements for partial-wave decomposition.  

The antiprotons of momenta between 1.5\,GeV/$c$ and 15\,GeV/$c$ delivered by the HESR together with the target protons at rest translate into centre-of-mass energies in the range of 2.2\,GeV $< E_{\rm cms} <$ 5.5\,GeV. The antiproton beam will impinge on a fixed target, being either of cluster-jet or frozen hydrogen pellet type for the $p\bar{p}$ annihilation programme. In addition, internal targets of heavier gases, such as deuterium, nitrogen or argon will be available for the $\bar{p}A$ studies. At a later stage, in particular for the planned hypernuclear experiments, non-gaseous nuclear targets will also be used, such as carbon fibres, thin wires or target foils.

The Micro Vertex Detector (MVD), surrounding the target region, will provide precise vertex position measurements of about 50\,$\mu$m perpendicular to and 100\,$\mu$m along the beam axis. It consists of silicon pixel and strip sensors. Tracking with a transverse momentum resolution d$p_{\rm T}/p_{\rm T}$ of better than 1\% will be provided by Gas Electro Multiplier (GEM) planes and a Straw Tube Tracker (STT) combined with the MVD and the field of the 2\,T solenoid magnet. Particle IDentification (PID) of pions, kaons and protons will be performed via two Detection of Internally Reflected Cherenkov Light (DIRC) detectors and a Time-Of-Flight detector system (TOF). State-of-the-art photon detection as well as separation between electrons and pions will be performed with an Electromagnetic Calorimeter (EMC) equipped with 17200 PbWO$_{\rm 4}$ crystals. Muon PID will be provided by the muon detector system surrounding the solenoid magnet.

The forward spectrometer covers polar angles below 10 and 5\,degrees in the horizontal and vertical plane, respectively. It comprises a forward tracking system (FTS) of three pairs of straw tube planes (each of the six stations equipped with four layers) before, inside and behind the 2\,Tm dipole magnet. An Aerogel Ring Imaging Cherenkov Counter (FRICH) and a Forward TOF system (FTOF) will be used for PID and the Forward Spectrometer Calorimeter (FSC) provides photon detection and electron/pion separation. A Forward Range System (FRS) and the Luminosity Detector (LMD) complete the forward spectrometer.

Since event rates of up to 20\,MHz are expected, the PANDA experiment will feature a triggerless readout. Reactions of interest will be selected online by complex algorithms running on compute nodes during data acquisition in order to maximise the fraction of events of interest to be recorded. Especially, the exemplary channel under study in this paper, {\it i.e.} the signature of a $J/\psi$ decaying to a high momentum lepton pair, benefits from this online filtering system. Significant background suppression at even highest expected event rates will be achieved.

\subsection{Resonance energy scans with PANDA/HESR}
\label{subsec:HESR}
The procedure of measuring the energy-dependent cross-section of a specific process over a certain range of centre-of-mass energies by adjusting (at high precision) the beam momentum is called a resonance energy scan.
\begin{figure}[tp!]
    \begin{center}
\includegraphics*[clip,trim= 5 2 5 5,width=0.3\textwidth,angle=0]{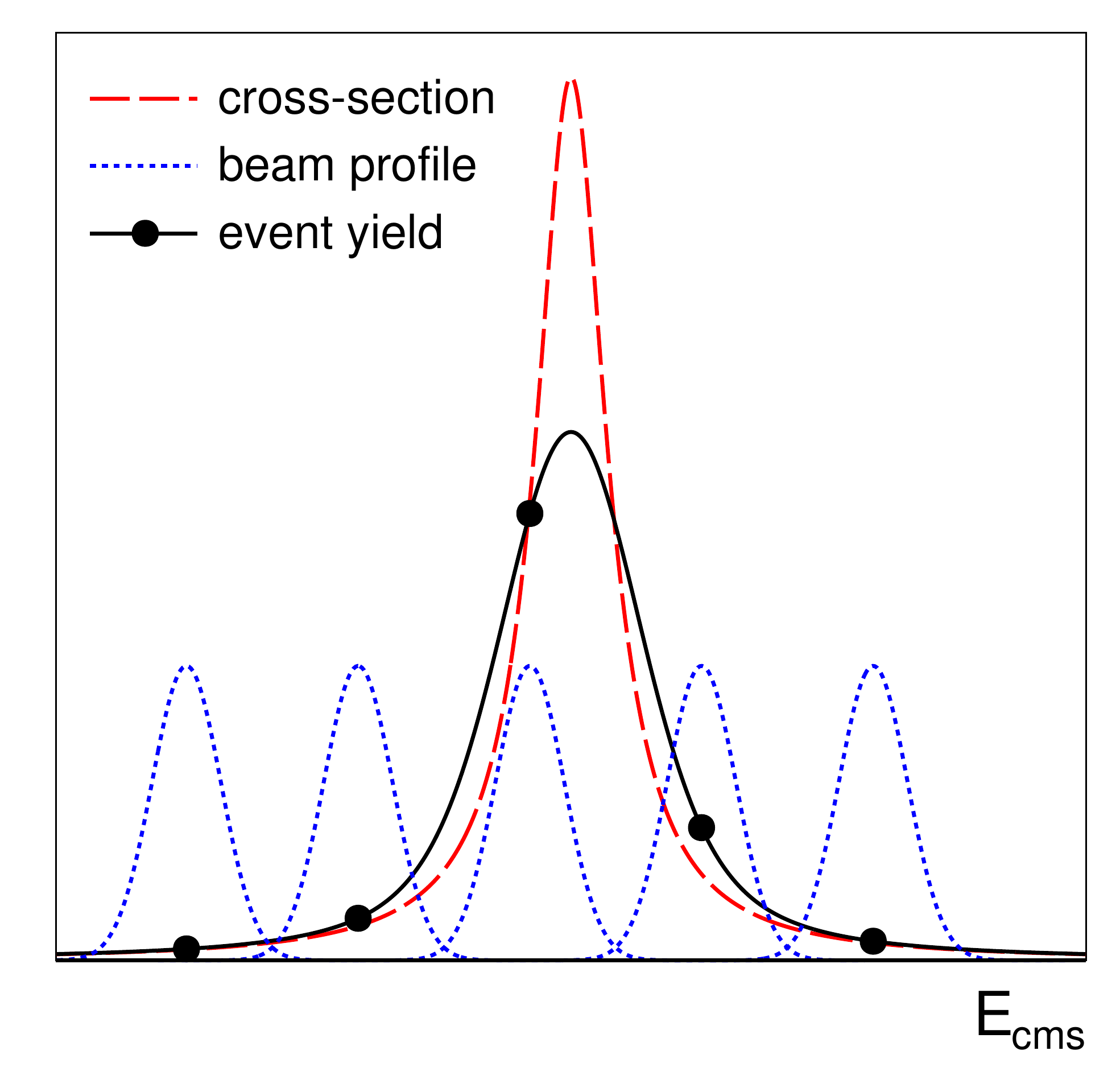}
    \end{center}
\vspace{-0.3cm}
      \caption{Schematic of a resonance energy scan. The true energy-dependent cross-section (dashed red line), 
        the beam energy profile (dotted blue line), the measured event yields (markers), and the effective 
        measured energy-dependent event yield (solid black line) are illustrated.}
\label{fig:scanscheme1} 
\end{figure}
Such scan procedure is illustrated in Fig.\,\ref{fig:scanscheme1}. The true energy-depen{-}dent cross-section is determined from the experimentally observed event yields at each centre-of-mass energy, by unfolding the (precisely known) beam energy profile.

Driving parameters for the resultant sensitivity performance for such a measurement are the number of reconstructed events per scan point, and thus the integrated luminosity $\cal{L}$, and the beam momentum spread d$p/p$, at which the antiprotons will be delivered by the HESR. We consider three different scenarios as they are expected for the different phases of the accelerator completion, see Tab.\,\ref{tab:Lumis}, extracted and computed based on~\cite{Lehrach2006}. Apart from the High Luminosity (HL) mode of the final accelerator setup with expected d$p/p=1 \cdot 10^{-4}$ and $\cal{L} \approx$ 13680\,(day $\cdot$ nb)$^{-1}$, there will be the High Resolution (HR) mode with a d$p/p$ about a factor of five more precise and an integrated luminosity about a factor of ten lower. 
\begin{table}[b]
\caption{Momentum spreads d$p/p$, beam-energy resolutions d$E_{\rm cms}$ and integrated luminosities $\cal{L}$ (at $\sqrt{s} = 3.872$ GeV) of 
  the three different HESR operation modes~\cite{Lehrach2006}.}
\vspace{1ex}
\label{tab:Lumis}
\centering
\begin{tabular}{c|c|c|c}
HESR mode & d$p/p$ & d$E_{\rm cms}$ [keV]& $\cal{L}$ [1/(day $\cdot$ nb)]\\\hline
 HL & $1 \cdot 10^{-4}$ & 167.8 & 13680 \\\hline
 HR & $2 \cdot 10^{-5}$ &  33.6 &  1368 \\\hline
 P1 & $5 \cdot 10^{-5}$ & 83.9  &  1170  \\\hline 
\end{tabular}
\end{table}
For the initial ``Phase-1'' (P1), the d$p/p$ is expected to be about another factor of two worse and the integrated luminosity lower by about 15\% than for the HR mode~\cite{PrasuhnSep15}. These three different HESR operation mode scenarios in terms of beam resolution and integrated luminosity are assumed in this study and the resonance energy scans are performed for each of them, respectively.

For the scan procedure itself, two accuracies are important. An initially set beam momentum can (after beam calibration) be determined with $10^{-4}$ relative uncertainty. The accuracy in relative beam adjustment (by frequency sweeping) will be $10^{-6}$, which is also the accuracy expected for the accelerator to reproduce a certain beam-momentum setting~\cite{LehrachPrashuhn}. Both uncertainties are taken into account accordingly as described in Sec.\,\ref{sec:SimuEscan}.

\subsection{Objectives of performed sensitivity studies}
\label{subsec:Objectives}
\begin{figure}[tp]
    \begin{center}
\includegraphics[clip, trim= 0 10 0 0, width=0.55\linewidth,angle=0]{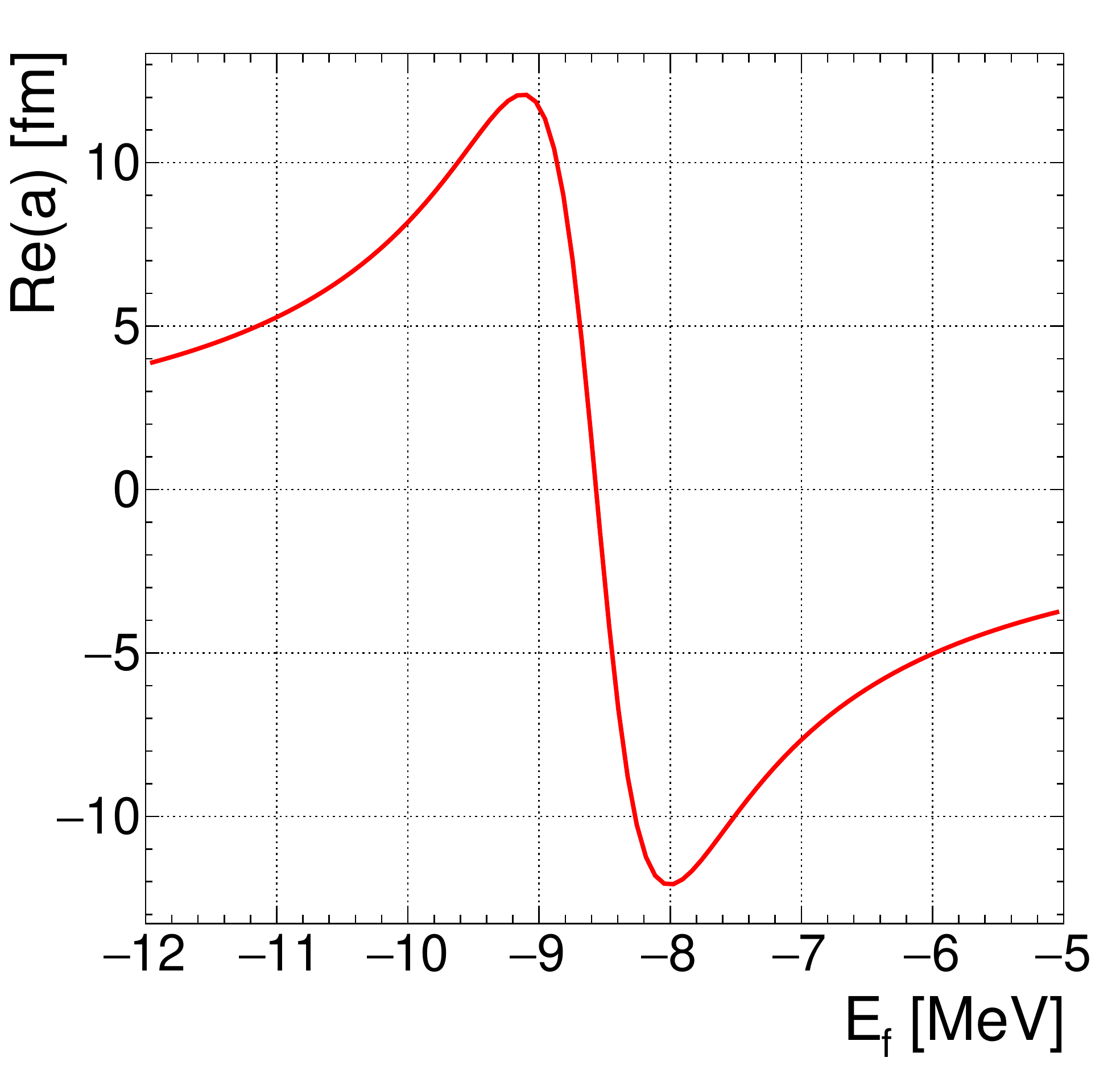}
    \end{center}
  \begin{picture}(1,1)
   \end{picture}
\vspace{-0.6cm}
      \caption{Functional dependence of the real part of the scattering length $\Re(a)$ on the parameter $E_{\rm f}$. 
        The condition $\Re(a)=0$ defines the threshold energy $E_{\rm f,th} \approx -8.5652$\,MeV, separating between 
        the $X(3872)$ being a bound ($\Re(a)>0$) or virtual ($\Re(a)<0$) state~\cite{HanhartLS,KalashLS}.}
\label{fig:Ef} 
\end{figure}
One goal of the energy-scan simulations is to study the achievable sensitivity for a natural decay width ($\Gamma_0$) measurement with PANDA for different HESR operation 
\linebreak 
modes in a realistic and limited data-taking time. The parameter $\Gamma_0$ is determined by fitting a Voigt function, a convolution of a Breit-Wigner function with a width $\Gamma_0$ and a Gaussian with a standard deviation $\sigma_{\rm Beam}$, accounting for the spread in beam momentum. This study is referred to as the Breit-Wigner Case in the following. The work presented here for this case of a width measurement extends and 
superceeds an initial preliminary study~\cite{Galuska}, in which just one input width for one assumed parameter set was addressed.

The narrow $X(3872)$ state used as an example here, is among others debated to be a loosely bound 
$(D\bar{D}^{\ast})^{0}$ molecule or a virtual scattering state effectively created by threshold dynamics~\cite{Hanhart_MoleculeReview}. As a consequence, the line shape would in such a scenario differ significantly from that of a 
simple Breit-Wigner-like resonance shape. It depends on the given decay channel (here $J/\psi\,\pi^+\pi^-$) and on the dynamic Flatt\'{e} parameter $E_{\rm f}$~\cite{HanhartLS,KalashLS} (corresponding to the inverse scattering length $\gamma$ in~\cite{BraatenLS}) that determines the nature of being either a bound or a virtual state. Therefore, we secondly study the sensitivity to distinguish between the two natures via the key parameter $E_{\rm f}$ (Fig.\,\ref{fig:Ef}). This study is referred to as the Molecule Case in the following.

\begin{figure}[tp!]
    \begin{center}
\includegraphics[clip, trim = 5 5 10 5,  width=0.48\linewidth,angle=0]{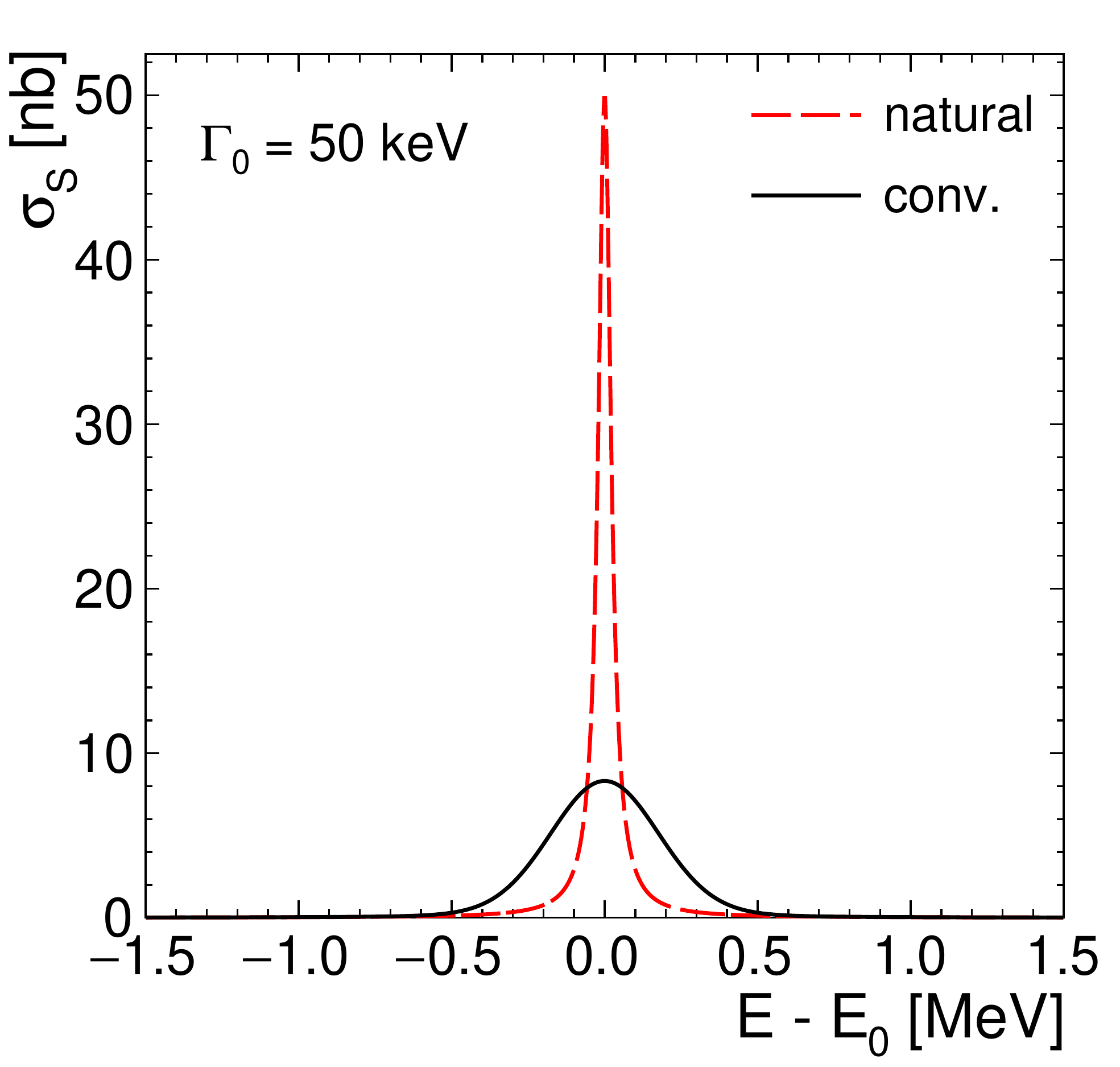}
\includegraphics[clip, trim = 5 5 10 5,  width=0.48\linewidth,angle=0]{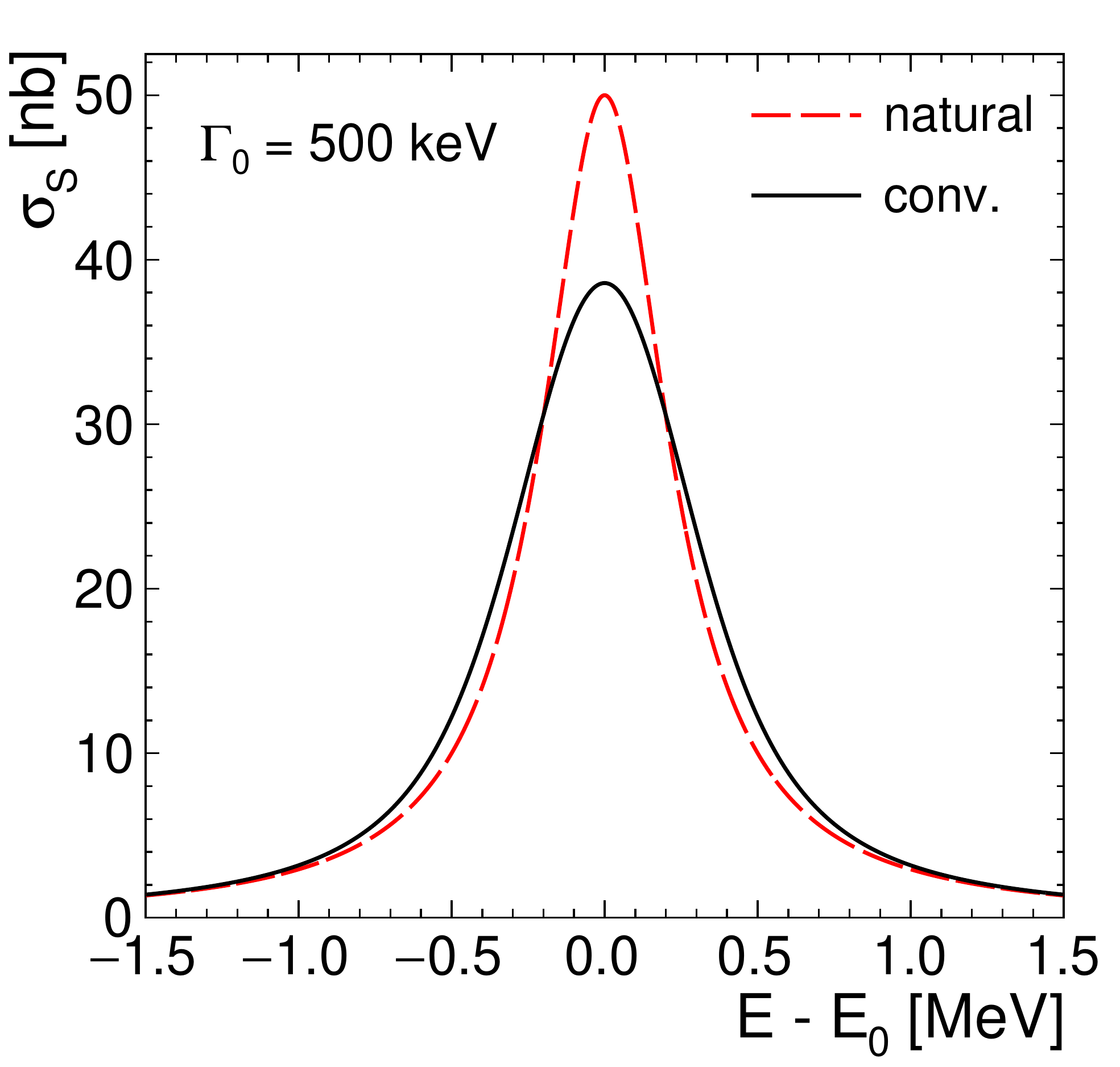}
\includegraphics[clip, trim = 5 5 10 5,  width=0.48\linewidth,angle=0]{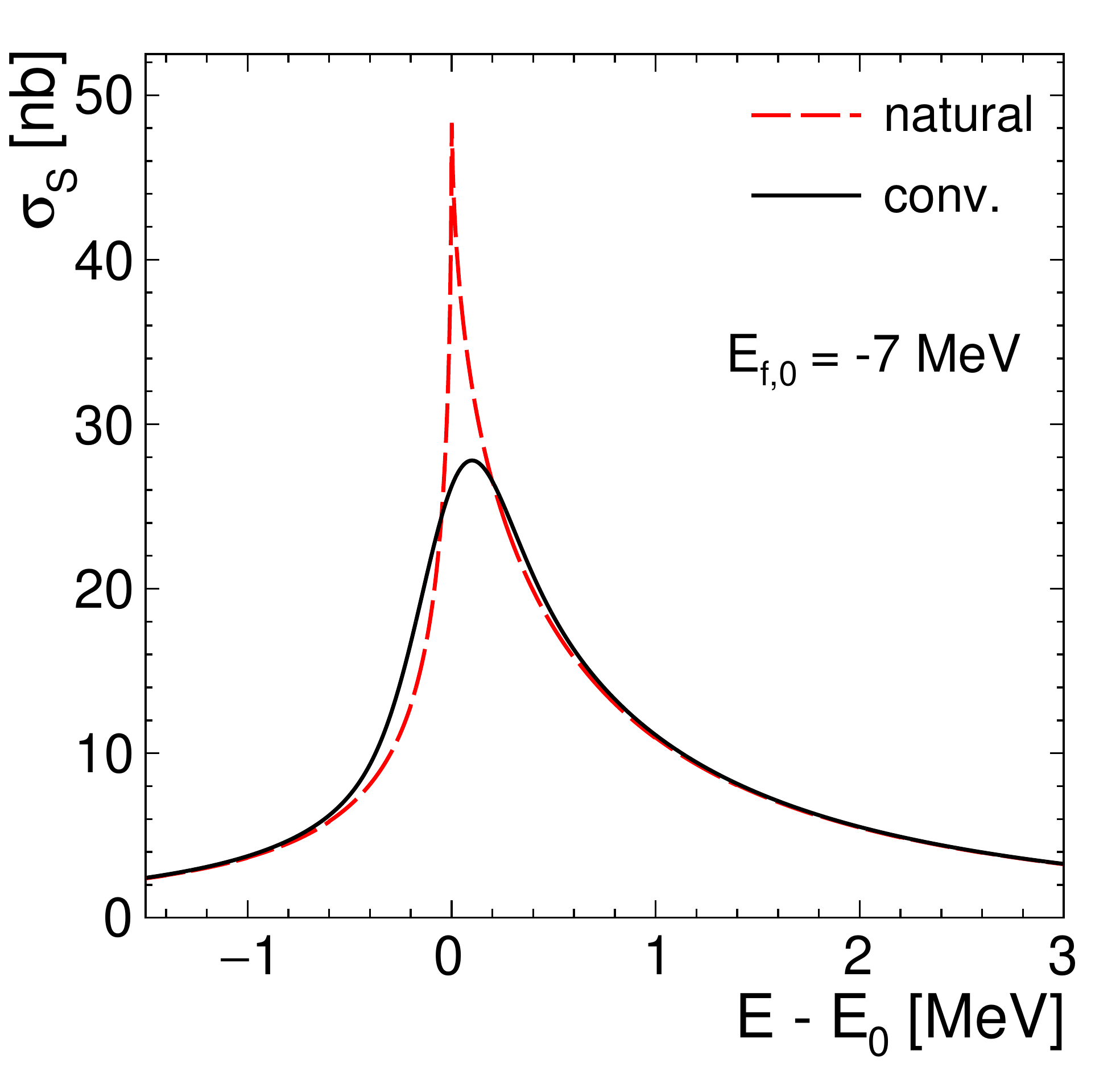}
\includegraphics[clip, trim = 5 5 10 5,  width=0.48\linewidth,angle=0]{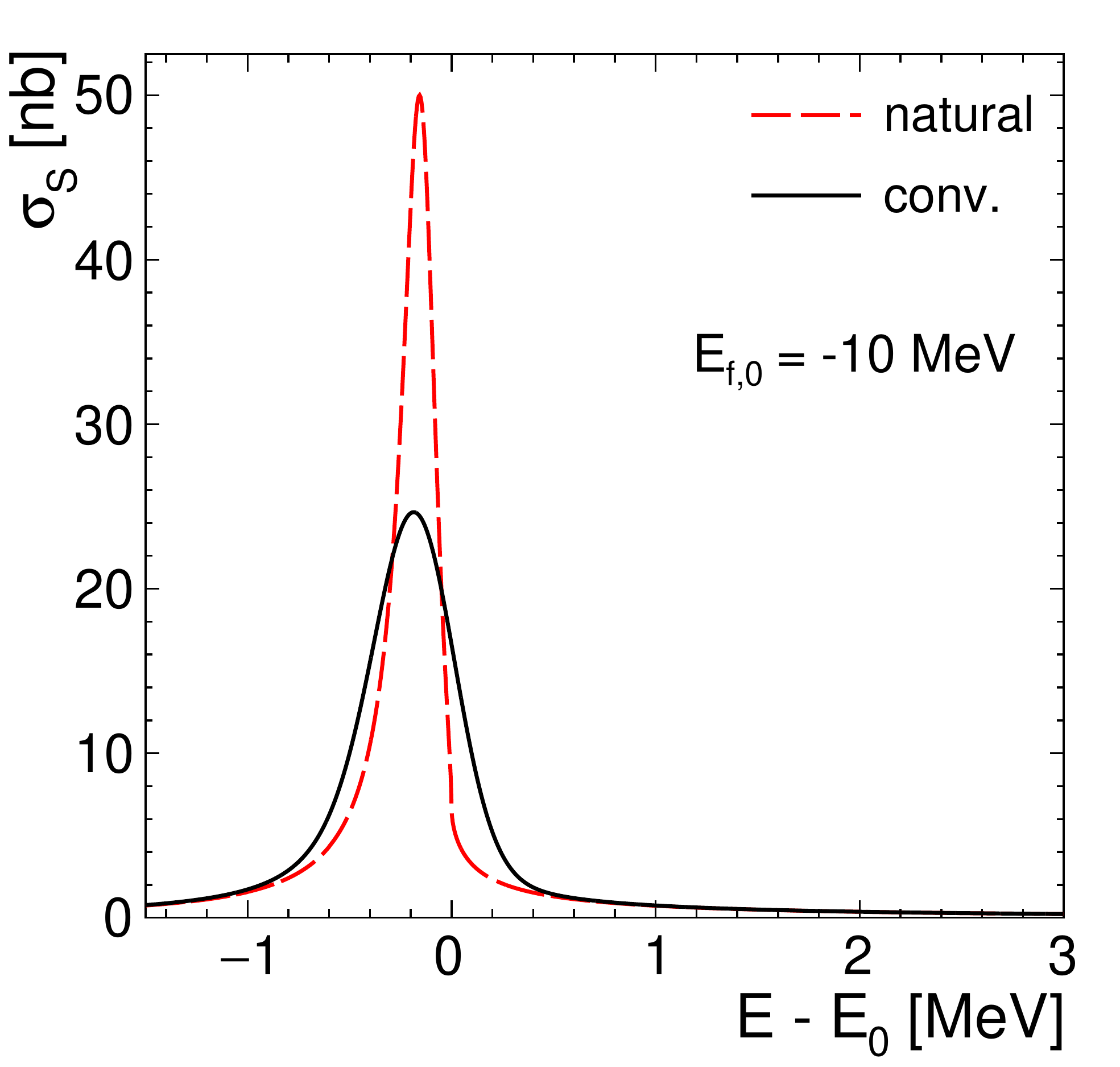}
  \begin{picture}(1,1)
 \put(-227,203){(a)}
 \put(-102,203){(b)}
 \put(-230,99){(c)}
 \put(-105,99){(d)}
  \end{picture}
    \end{center}
\vspace{-0.3cm}
      \caption[]{True physical (dashed red) and beam-resolution convoluted line shapes (solid black) for different examples of 
        Breit-Wigner widths $\Gamma_0$ (a, b) and $E_{\rm f,0}$ parameters for the Molecule Case (c, d), shown here for $\sigma_{\rm S,0}=50$\,nb.}
\label{fig:xlineshapes} 
\end{figure}
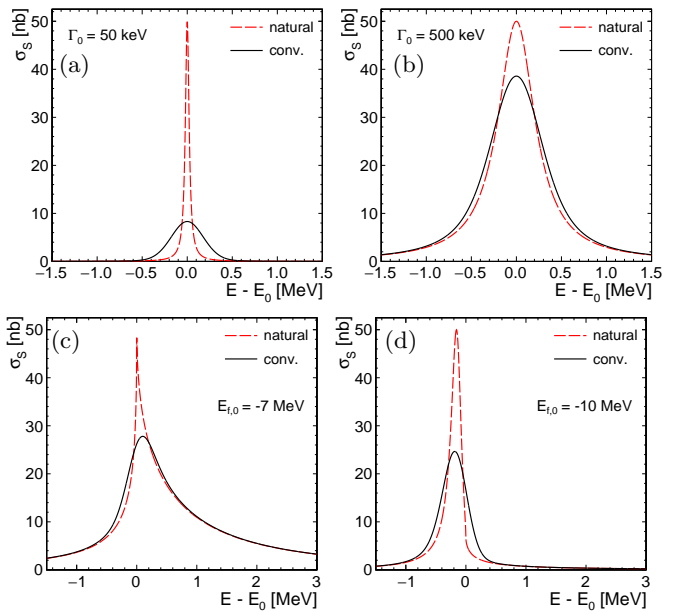

Examples of true physical and beam-resolution convoluted line shapes are shown for illustration for both, Breit-Wigner and Molecule 
Cases in Fig.\,\ref{fig:xlineshapes}.

\section{Event simulation and reconstruction}
\label{sec:EventSelection}
All Monte Carlo (MC) event simulations and reconstruction have been performed within the Geant-based 
\linebreak 
{\tt PandaRoot} software framework~\cite{PandaRoot}. The generated and simulated MC data are reconstructed to extract the reconstruction efficiencies for signal and for background events, as needed as input for the realistic simulation of the resonance energy scan (Sec.\,\ref{sec:SimuEscan}). Since the energy-scan window only ranges up to a few MeV, we considered a constant reconstruction efficiency for the different energy points $E_{{\rm cms,} i}$ of a given scan range. The MC signal and background data are all generated at $E_{\rm cms}= 3872\,$MeV.

\subsection{Monte Carlo event generation}
\label{subsec:MCsim}
\paragraph{Signal events}
\begin{table}[b]
\caption{Number of simulated signal and background events.}
\vspace{1ex}
\label{tab:data}
\centering
\renewcommand{\arraystretch}{1.1}
\begin{tabular}{c|l|r}
Type & Description & Generated events\\\hline
\multirow{2}*{S} 
& $\bar{p}p\rightarrow J/\psi \,\rho^0\rightarrow e^+ e^- \pi^+\pi^-$ & 98\,k\\
&$\bar{p}p\rightarrow J/\psi \,\rho^0\rightarrow \mu^+ \mu^- \pi^+\pi^-$ &  100\,k\\\hline
\multirow{2}*{NR} 
&$\bar{p}p\rightarrow J/\psi(\rightarrow e^+ e^-) \,\pi^+\pi^-$ & 100\,k\\
&$\bar{p}p\rightarrow J/\psi(\rightarrow \mu^+ \mu^-) \,\pi^+\pi^-$& 99\,k\\\hline
\multirow{2}*{gen} 
&DPM ($J/\psi\rightarrow e^+e^-$) & 9.6\,B gen. / 10\,M sim. \\
&DPM ($J/\psi\rightarrow \mu^+\mu^-$) & 8.9\,B gen. / 10\,M sim.\\\bottomrule 
\end{tabular}
\end{table}
\renewcommand{\arraystretch}{1.}
The reactions of interest are those with an intermediate $X(3872)$ produced in formation, $\bar{p}p \to X(3872)$. It is reconstructed in the decay channels
\begin{eqnarray}
X(3872)&\rightarrow &J/\psi \,\rho^0 \rightarrow e^+ e^- \pi^+ \pi^-~\mbox{and}\\
X(3872)&\rightarrow &J/\psi \,\rho^0 \rightarrow \mu^+ \mu^- \pi^+ \pi^-~,
\label{eq:decay}
\end{eqnarray}
which both offer a good suppression of light hadronic background due to the particular kinematics of the rather heavy $J/\psi$ decaying into two light leptons. The signal events ($S$) with the subsequent decay $J/\psi\rightarrow l^+l^-$ have been generated with {\tt EvtGen}~\cite{EvtGen}, see Tab.\,\ref{tab:data}.
\begin{figure*}[tp!]
    \begin{center}
\includegraphics*[clip,trim= 1 1 5 5,width=0.84\textwidth,angle=0]{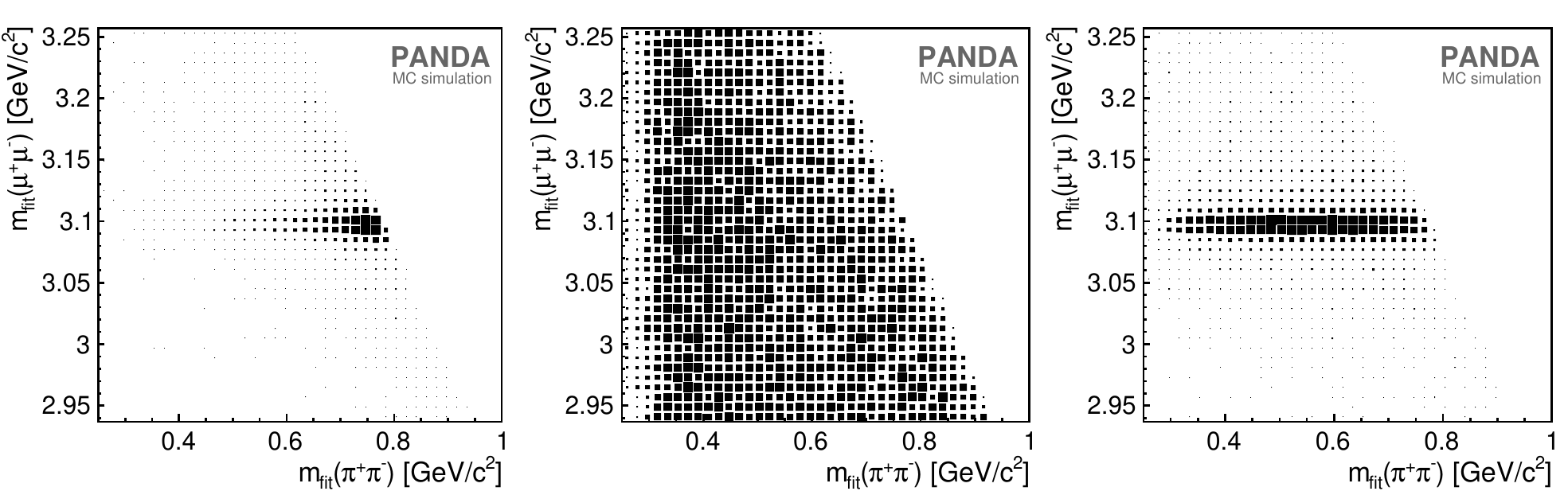}
    \end{center}
\vspace{-0.15cm}
      \caption{Di-muon {\it vs.} di-pion invariant mass spectra after a kinematic 4-constraint fit for the three different event types. 
        Two-dimensional distributions of the reconstructed events in $m_{\rm fit}(\mu^+ \mu^-)$ vs. $m_{\rm fit}(\pi^+\pi^-)$ for signal 
        events {\it (left)}, generic hadronic background {\it (centre)} and non-resonant $J/\psi$ background without intermediate 
        $\rho^0${\it (right)} after pre-selection. The correlation between the variables is clearly visible.}
\label{fig:MCsamplesMuCase_AfterPreSelect} 
\end{figure*}

\paragraph{Non-resonant  \boldmath{$J/\psi$} background}
There is no simple possibility to distinguish between reactions $\bar{p}p\rightarrow X(3872) \rightarrow J/\psi \,\pi^+\pi^-$ (with intermediate resonant formation) and 
\linebreak 
$\bar{p}p \rightarrow J/\psi \,\pi^+\pi^-$ (non-resonant). This is because the total centre-of-mass energy $E_{\rm cms}$ is the same in the resonant and non-resonant (NR) decay mode. The distribution of the invariant mass of the reconstructed exclusive system provides no more information than the total energy resolution introduced by the reconstruction based on the detectors. The mass of the $J/\psi$ subsystem in each case should therefore be essentially identical, this correlation can directly be seen in Fig.\,\ref{fig:MCsamplesMuCase_AfterPreSelect}. As a consequence, this kind of background must be taken into account. Even if being produced with a rather low cross-section, it cannot completely be separated from true signal events. Given the signal events would always comprise an intermediate $\rho^0(770)$ (as assumed here), a handle for separation is given by the different kinematics for $\rho^0(770) \rightarrow \pi^+\pi^-$ as compared to the non-resonant $\pi^+ \pi^-$ production. 

In order to investigate the impact, non-resonant $J/\psi$ background events have accordingly been generated, simulated and reconstructed (Tab.\,\ref{tab:data}). The effect of non-
\linebreak
resonant events that include an intermediate $\rho^0(770)$ is accounted for by a systematic error (Sec.\,\ref{subsec:Systs}). 
Interference effects between signal and non-resonant amplitudes have not been considered in this study. The assumed 
signal cross-sections can be considered as effective cross-sections, including interferences.

\paragraph{Generic hadronic background}
The total inelastic cross-section $\sigma_{\bar{p}p\rightarrow {\rm hadrons}}$ is about a factor $10^6$ larger than the assumed signal cross-sections (Tab.~\ref{tab:parasum}, Sec.\,\ref{subsec:ParamSpace}). Therefore, generic (gen) background reactions (predominantly producing light hadrons), although comprising different final state particles, can produce a significant contamination via the effect of missing, secondary as well as misidentified particles. 

The simulation of generic hadronic reactions is based on the Dual Parton Model (DPM) generator~\cite{DPM}. In order to achieve a signal-to-background ratio $S/B >1$, the background reconstruction efficiency $\epsilon_{B}$ needs to be limited by
\begin{eqnarray}
\frac{S}{B} &=& \frac{\sigma_{S} \cdot \epsilon_{S}}{\sigma_{B} \cdot \epsilon_{B}} \cdot {\cal B}_{J/\psi}\cdot {\cal B}_{X} > 1 \\
\Rightarrow \epsilon_{B} &<& 6.5 \cdot 10^{-10}~,\nonumber
\label{eq:ston_a}
\end{eqnarray}
when one assumes reasonable numbers (Tab.~\ref{tab:parasum}, Sec.\,\ref{subsec:ParamSpace}) for the signal and background production cross-sections of $\sigma_{S,{\rm 0}}=50$\,nb and $\sigma_{B} = 46$\,mb, a signal reconstruction efficiency of about $\epsilon_S = 10$\%, branching fractions ${\cal B}_{J/\psi}$ $:= {\cal B}(J/\psi\rightarrow \ell^+\ell^-) = 12$\%, and ${\cal B}_X$ $:= {\cal B}(X(3872)\rightarrow J/\psi \,\pi^+\pi^-)=5$\% as concluded from~\cite{PbarXCERN,LHCbXpp,LHCbXppUpdate,BelleXJ2pi}. To measure $\epsilon_{B}$ with at least one residual reconstructed background event $B_{\rm rec} \geq 1$, the minimum number of simulated generic background events $B_{0,{\rm gen}}$ can thus be estimated as
\begin{eqnarray}
\Rightarrow B_{0,{\rm gen}} &=& \frac{B_{\rm rec}}{\epsilon_B} > \frac{1}{6.5\cdot 10^{-10}} = 1.5\cdot 10^9~.
\label{eq:ston}
\end{eqnarray}

The above calculation shows the number of events 
\linebreak
needed is larger than one billion and implies in addition, that all except (at least) one background events are rejected, putting a quite demanding requirement to the selection process. To reduce the CPU-wise effort, we apply a pre-filter already at the generator level selecting only those events with at least four tracks $t$, out of which two tracks with opposite charge create an invariant mass within [$2.8 < m(t^+t^-) < 3.3]$\,GeV/$c^2$ as input for the Geant-based MC simulation and further reconstruction. This saves the computing time spent for simulating events with the di-lepton mass incompatible with a $J/\psi$. The number of events that effectively needs to be simulated and reconstructed is hereby reduced by about a factor of 1000, {\it i.e.} the $10^7$ simulated and reconstructed generic background events correspond to about $10^{10}$ generated events as quoted in Tab.\,\ref{tab:data}, see also Fig.\,\ref{fig:DiLeptonSpectra_AfterFinalSelect}.

\subsection{Event selection}
\label{subsec:EvtSelection}
All four-particle combinations of two oppositely charged lepton and two oppositely charged pion candidates are reconstructed, combined and a four-constraint (4C) kinematic fit to the initial $\bar{p}p$ system is performed. For the lepton candidates, the PID probabilities $P_{\rm PID}$ (based on all PANDA PID detectors) are required to fulfil $P_{\rm PID}(e) > 0.95$ and $P_{\rm PID}(\mu) > 0.99$, and a cut on the resultant $\chi^2_{\rm 4C}(e^+e^-\pi^+\pi^-) < 200$ and $\chi^2_{\rm 4C}(\mu^+\mu^-\pi^+\pi^-)< 100$ is applied, respectively. In order to optimise the selection, further cuts have been studied and applied. The $J/\psi \to e^+e^-$ decay is selected by a cut on the invariant mass of $|m_{\rm fit}(e^+e^-) - 3.095$\,GeV/$c^2| < 0.2$\,\,GeV/$c^2$ and $m_{\rm fit}(e^+e^-)$ $+ m_{\rm fit}(\pi^+\pi^-)$ $> 3.77$\,GeV/$c^2$ is requested for the sum of the two invariant masses, the latter implicitely selects the $\rho^0(770)$. The four-particle momentum is restricted to 
\linebreak 
$p_{\rm cms}(e^+e^- \pi^+\pi^-) < 0.4$ GeV/$c$, and the di-lepton opening angle is requested to be $\angle(p_{e^+}, p_{e^-})< 2.1$\,rad. 
\begin{figure}[tp!]
     \begin{center}
       \includegraphics[clip,trim= 2 1 5 5,width=1.0\linewidth, angle=0]{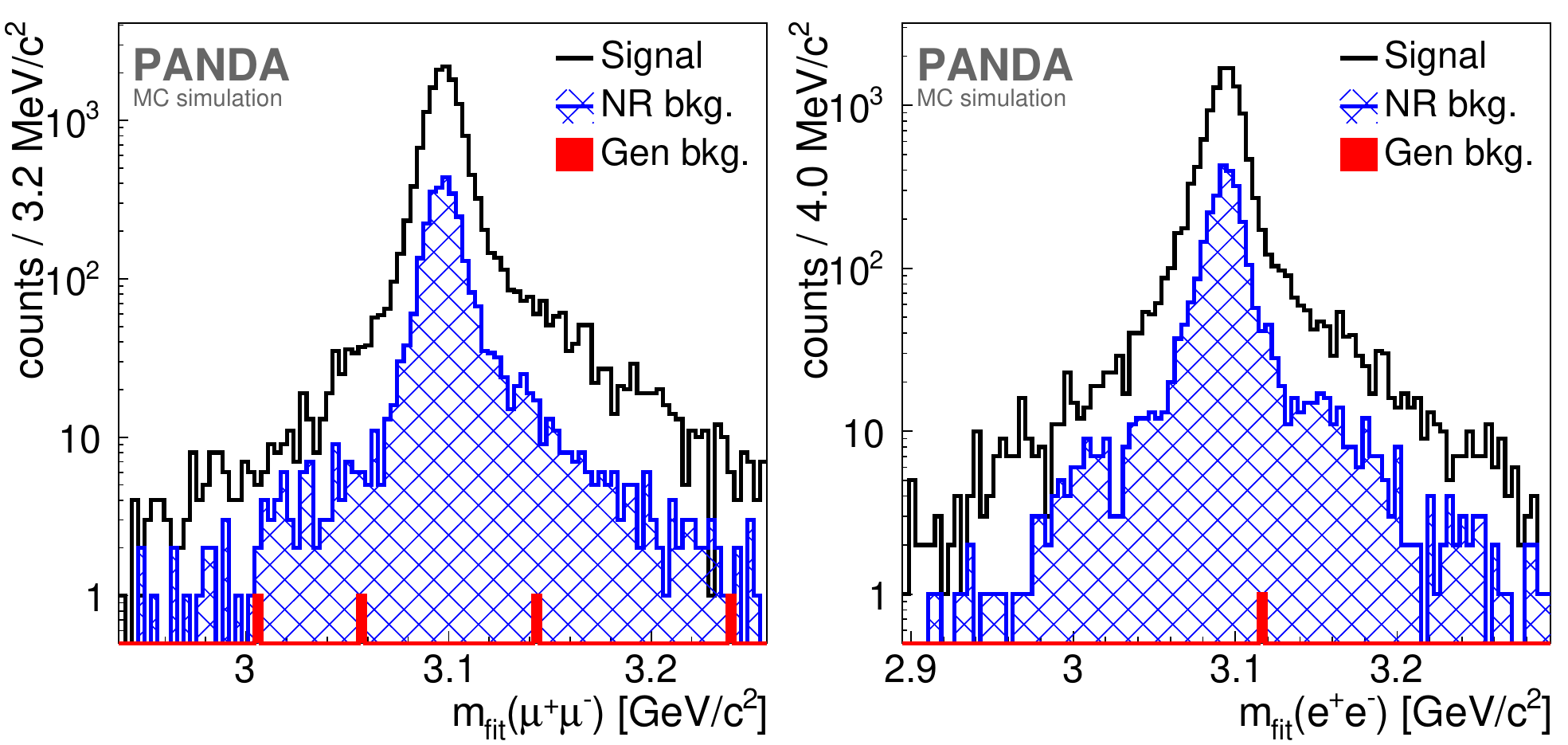}
     \end{center}
\vspace{-0.15cm}
      \caption{Reconstructed invariant di-lepton mass distribution for the muon {\it (left)} and electron {\it (right)} channel 
        for signal {\it (black)}, non-resonant (blue) and generic DPM background {\it (red)} MC data, after final selection.
        The mass ranges shown have been chosen to be about $\pm 10\sigma$ around the $J/\psi$ and define the regions of interest 
        for the maximum-likelihood fits performed (Sec.\,\ref{sec:SimuEscan}). 
      }
\label{fig:DiLeptonSpectra_AfterFinalSelect} 
\end{figure}
For the $J/\psi \to \mu^+\mu^-$ channel, the $J/\psi$ mass window cut $|m_{\rm fit}(\mu^+\mu^-) - 3.097$\,GeV/$c^2| < 0.16$\,\,GeV/$c^2$, the mass sum cut $m_{\rm fit}(\mu^+\mu^-) + m_{\rm fit}(\pi^+\pi^-) > 3.78$\,GeV/$c^2$ and the one on the di-lepton opening angle of $\angle(p_{\mu^+}, p_{\mu^-}) < 2.1$\,rad are similarly applied. In the muon case, an additional cut on the event sphericity~\cite{Sphericity} requested to be $< 0.11$ further improves $S/B$. 
\renewcommand{\arraystretch}{1.1}
\begin{table}[b]
\small
\caption{Reconstruction efficiencies as obtained from Geant-based detector MC simulation. 
}
\vspace{2ex}
\label{tab:effi}
\centering
\begin{tabular}{c|c|c|c} 
 $X(3872)\rightarrow J/\psi \pi^+\pi^-$        & $\epsilon_S$ & $\epsilon_{B,{\rm gen}}$ & $\epsilon_{B,{\rm NR}}$ \\\hline
 $J/\psi\rightarrow e^+e^-$         &  12.2\%  & 1.0  $\cdot 10^{-10}$   &  2.8\%   \\
 $J/\psi \rightarrow \mu^+\mu^-$    &  15.2\%   &  4.5 $\cdot 10^{-10}$ &  3.0\%    \\
\end{tabular}
\end{table}
\renewcommand{\arraystretch}{1.}

The reconstruction efficiencies after event selection are summarised in Tab.\,\ref{tab:effi}. They are significantly lower for the $e^+e^-$ channel due to final-state radiation not being completely recovered in the 4C kinematic fit. In order to determine the $J/\psi$ yields needed for our approach of line shape measurements (Sec.\,\ref{sec:SimuEscan}), the $J/\psi$ mass window cuts on the di-lepton masses are not tightened, they correspond to about $\pm 10\sigma$ for both channels (Fig.\,\ref{fig:DiLeptonSpectra_AfterFinalSelect}).

\section{Simulation of resonance energy scans}
\label{sec:SimuEscan}
For simulating a resonance energy scan, the corresponding reconstruction efficiency $\epsilon_i$, the integrated luminosity at each scan point $E_{{\rm cms},i}$ and the effective cross-section $\sigma^*(E_{{\rm cms},i})$ are needed as input. The expected signal yield is determined from
\begin{linenomath*}
\begin{equation}
\label{Eq.SigYield}
N_i = \sigma^*(E_{{\rm cms},i}) \cdot {\epsilon_{i} \cdot {\cal L}_{i} \cdot f_{\rm {\cal B}}}~,
\end{equation}
\end{linenomath*}
where ${\cal L}_{i}$ is the integrated luminosity accumulated at the $i^{\rm th}$ energy scan point $E_{{\rm cms}, i}$ and $f_{\rm {\cal B}}$ stands for the product of all involved branching fractions of the reconstructed decay. The observable line shape is given by the convolution of the original line shape $\sigma(E;P)$ with the resolution function $G(E;\sigma_{\rm beam})(E)$
\begin{linenomath*}
\begin{equation}
\sigma^{*}(E;P) = \int_0^{\infty}\, \sigma(E';P) \cdot G(E-E';\sigma_{\rm beam})\, dE'~.
\end{equation}
\end{linenomath*}
Together with an appropriate background description function $B(E)$, one can fit this function to the final energy-dependent distribution to extract the absolute cross-sec{-}tion $\sigma_{S}$ and the parameters $P$ of interest. In case one is not interested in the absolute cross-section, this functional shape can be fitted directly to the measured (here simulated) event yields $N_{i}$ to extract the (resonance) parameters, {\it cf.} Fig.\,\ref{fig:scanscheme1}. 

For our study, a (random) jitter of $10^{-4}$ is applied as uncertainty for setting the first energy scan point, whereas a relative uncertainty of $10^{-6}$ is applied when setting all further energy scan points relatively to the previous one. This is done for each performed energy scan measurement, respectively, taking into account the nominal HESR specifications, {\it cf.} Sec.\,\ref{subsec:HESR}. 

\subsection{Physics parameter space}
\label{subsec:ParamSpace}
\renewcommand{\arraystretch}{1.1}
\begin{table}[bp]
\small
\caption{Summary of parameter settings under consideration.}
\vspace{1ex}
\label{tab:parasum}
\centering
\begin{tabular}{c|c}
Input parameter & Assumed value(s) \\\hline
${\cal B}(X\rightarrow J/\psi\,\rho^0)$       & 5\% ~\cite{BelleXJ2pi,BES3BRX,CDFJrho}\\
${\cal B}(J/\psi\rightarrow e^+e^-)$      & 5.971\%  ~\cite{PDG15}\\
${\cal B}(J/\psi\rightarrow \mu^+\mu^-)$     & 5.961\% ~ \cite{PDG15}\\
${\cal B}(\rho^0\rightarrow \pi^+\pi^-)$    &  100\% ~\cite{PDG15}\\\hline
\multirow{2}*{$\sigma_{S}$}    &  50\,nb~\cite{LHCbXpp,LHCbXppUpdate,BelleXJ2pi}\\
                                  &  [20, 30, 75, 100, 150]\,nb\\ 
$\sigma_{B,\mbox{\scriptsize gen}}$  & 46\,mb~\cite{PbarXCERN} \\
$\sigma_{B,\mbox{\scriptsize NR}}$      & 1.2\,nb~ \cite{ChenNRBG}\\\hline 
Total scan time $t_{\rm scan}$   & 80\,d\\
No of scan points $N_{\rm scan}$ & 40  \\\hline
\multirow{2}*{Breit-Wigner, $\Gamma_X$}    & $[50,70,100,130,~~~~$  \\
                                                  & $~~~~180,250,500]$\,keV  \\
\multirow{2}*{Line shape, $E_{\rm f}$} & $-[10.0, 9.5, 9.0, 8.8,~~~~$ \\
                                       & ~~~~~$~~~8.3, 8.0, 7.5, 7.0]$\,MeV \\\bottomrule
\end{tabular}
\end{table}
\renewcommand{\arraystretch}{1.1}
The simulation of a scan measurement requires assumptions on unknown physics quantities, such as the signal production cross-sections $\sigma_{S}$ and natural decay widths $\Gamma_{\rm 0}$ or line shape parameters $E_{\rm f}$. Based on the foreseen HESR parameters (Sec.\,\ref{subsec:HESR}), a multidimensional parameter space of possible scenarios is investigated in this sensitivity study. A summary of the physics parameters that were considered in the analyses are given in Tab.~\ref{tab:parasum}, where applicable together with the relevant references.

Since we use the example of the directly in formation produced $X(3872)\rightarrow J/\psi\,\rho^0(770)$, the corresponding branching fraction is essential for the expected number of reconstructed events. To date, this branching fraction is experimentally restricted to $2.6\% < {\cal B}(X\rightarrow J/\psi\,\pi^+\pi^-) < 6.6\%$~\cite{BelleXJ2pi}. Following~\cite{BES3BRX} and assuming the decay populates this final state predominantly via $J/\psi\,\rho^0(770)$, as suggested by the observed di-pion spectra \cite{CDFJrho}, a value of ${\cal B}(X\rightarrow J/\psi\,\rho^0(770))\equiv {\cal B}(X\rightarrow J/\psi\,\pi^+\pi^-) =5\%$ is assumed for this study, corresponding to about the centre of the experimentally restricted range.
 
In order to estimate the number of signal events for a given integrated luminosity, an assumption about the peak production cross-section $\sigma_{\rm \bar{p}p\rightarrow X(3872)}:=\sigma_{S,{\rm 0}}$ is required. Since no direct measurement of this number exists yet, we compute an estimate via crossing symmetry by using the expression for resonance formation assuming Breit-Wigner dynamics~\cite{PDG15}:
\begin{linenomath*}
\begin{align}
\label{Eq.CrSym}
&\sigma_{i\rightarrow R \rightarrow f}(E)\\\nonumber
&=\frac{(2J+1)}{(2S_1+1)(2S_2+1)}\frac{4\pi}{k^2}\left[\frac{\Gamma^2/4}{(E-E_R)^2+\Gamma^2/4}\right]
{{\cal B}_i\cdot {\cal B}_f}~,
\end{align}
\end{linenomath*}
where $i$ represents the initial state system, $f$ the final state system, $R$ the formed resonance (here the $X(3872)$:=$X$), $k^2 = (M_X^2-4m_p^2)/4$ the squared break-up momentum for the resonance mass $M_X$ and the proton mass $m_p$, and ${\cal B}$ the corresponding branching fractions of $R$ going to either $i$ and $f$. 
\begin{figure}[tp!]
  \centering
  \includegraphics[clip, trim= 395 0 0 27,width=0.75\linewidth,angle=0]{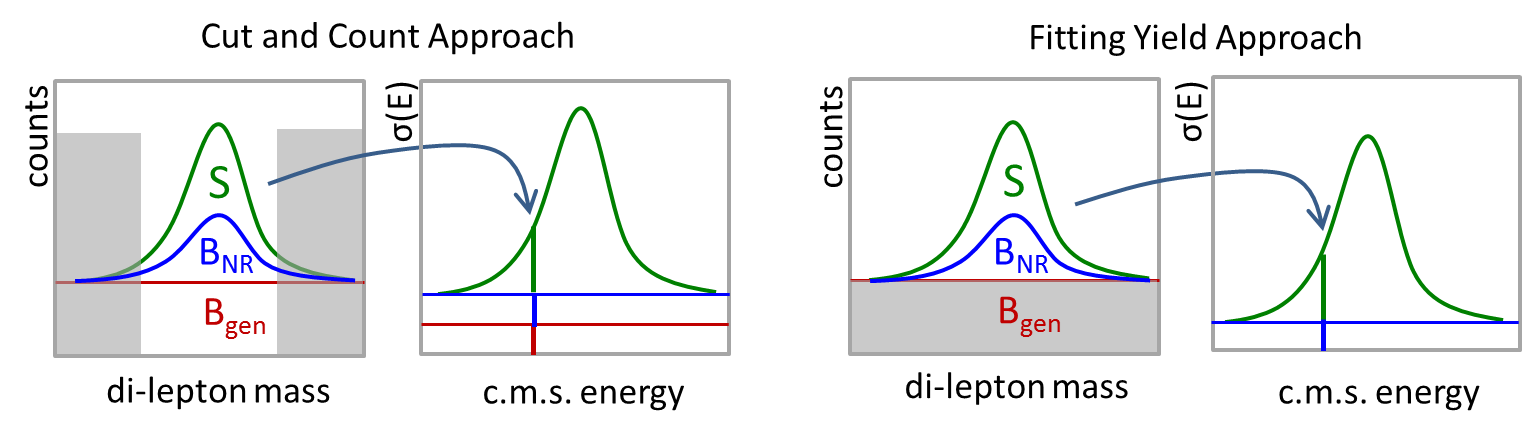}
      \caption{By fitting the yield of $J/\psi$ in the di-lepton candidate mass distribution {\it (left)}, 
        one is able to remove the flat generic DPM background $B_{\rm gen}$ (red) and keeps contributions from the 
        non-resonant backgrounds $B_{\rm NR}$ (blue) in form of a constant background contribution 
        in the result of the energy-dependent cross-section {\it (right)}.}
\label{fig:scantech} 
\end{figure}

For the initial state $\bar{p}p$, we evaluate Eq.\,\ref{Eq.CrSym} using the upper limits on ${\cal B}(X\rightarrow\bar{p}p)/{\cal B}(X\rightarrow J/\psi\,\pi^+\pi^-)<6.3\cdot 10^{-4}$ \cite{LHCbXppUpdate} and ${\cal B}(X\rightarrow J/\psi\,\pi^+\pi^-)<6.6\%$ 
at the resonance position $E = E_R = M_X c^2$ for the (proton) spins $S_1=S_2=1/2$ and $J_X=1$. We obtain an upper limit on the signal production cross-section $\sigma_{S} < 52.8$\,nb. Apart from this value of $\approx$ 50\,nb based on actual experimental measurements, we further assume smaller and larger input cross-sections to cover a larger range, and thus address achievable sensitivities also for other possible states of interest in the future.    

We have chosen a reasonable dedicated data taking time of in total $t_{\rm scan}=80$ days for one energy-scan measurement, equally shared for $N_{\rm scan}=40$ different center-of-mass energies, means two days of data taking per $E_{{\rm cms},i}$. The scan points are equidistantly distributed over a given energy scan range, resulting in scan point distances between about d$E_{\rm cms} \approx 30\,$keV and d$E_{\rm cms} \approx 200\,$keV, depending on the given input parameters and physics case. 

Given the experimental upper limit of 1.2\,MeV on the natural decay width of the $X(3872)$ (Sec.\,\ref{intro}), we have chosen seven input values for $\Gamma_{\rm 0}$ in the sub-MeV range, covering a few tens up to a few hundreds of keV for the Breit-Wigner Case. For the Molecule Case, the distinction of the nature becomes experimentally challenging for $E_{\rm f}$ parameters close to $E_{\rm f,th}$, wherefore we also chose here several input values $E_{\rm f,0}$ covering a sub-MeV range around $E_{\rm f,th}$. 

For each simulated scan experiment, we set the input values according to  Tab.\,\ref{tab:Lumis} and Tab.\,\ref{tab:parasum} and perform the full procedure, simulation and analysis as described in the following (Secs.\,\ref{subsec:SiumYields} - \ref{subsec:Systs}), to extract the sensitivity for each combination of the accelerator scenario (Tab.\,\ref{tab:Lumis}), the input signal cross-section and the physics input parameter $\Gamma_{\rm 0}$ or $E_{\rm f,0}$ for the Breit-Wigner and the Molecule Case, respectively (Tab.\,\ref{tab:parasum}).    
   
\subsection{Simulation and extraction of energy-dependent event yields}
\label{subsec:SiumYields}
In order to measure the energy-dependent cross-section, we need to determine the numbers of signal events at the different centre-of-mass energy positions $E_{{\rm cms},i}$ as described above. Since we do not have real data, we simulate these event yields, which in turn are extracted from the simulated data for our sensitivity studies. 

After event selection, the data is composed of signal, non-resonant background and generic background events. Since all these remaining events fulfill the 4C fit (Sec.\,\ref{sec:EventSelection}) they contribute to a peak in the $\ell^+\ell^-\pi^+\pi^-$ invariant mass spectrum. In the di-lepton invariant mass, however, the generic background is flat and can thus be separated 
\linebreak
(Fig.\,\ref{fig:scantech}, left). Therefore, we measure the $J/\psi$ signal peak content by a maximum-likelihood (ML) fit to the invariant di-lepton mass distribution that delivers the $\bar{p}p \rightarrow J/\psi\pi^+\pi^-$ event yield at a given scan point $E_{{\rm cms}, i}$. The resultant energy-dependent event yield distribution (Fig.\,\ref{fig:scantech}, right) still comprises some non-resonant background contribution that can be described by an additional constant in the fit function, while the rather flat generic background is not present anymore.

The following procedure is applied to simulate the expected event rates for further analysis. First, we compute the expected number of observed signal and background events according to Eq.\,\ref{Eq.SigYield}, where we use the reconstruction efficiencies determined by the Geant-based MC simulation and reconstruction (Fig.\,\ref{fig:DiLeptonSpectra_AfterFinalSelect}) described in Sec.\,\ref{sec:EventSelection} and listed in Tab.\,\ref{tab:effi}. We assume that all three signal and background efficiencies are constant, {\it i.e}. $\epsilon_i = \epsilon$, across the rather narrow energy scan window. 

\begin{figure}[tp]
\begin{center}
\includegraphics[clip, trim= 1 0 0 0, width=0.99\linewidth,angle=0]{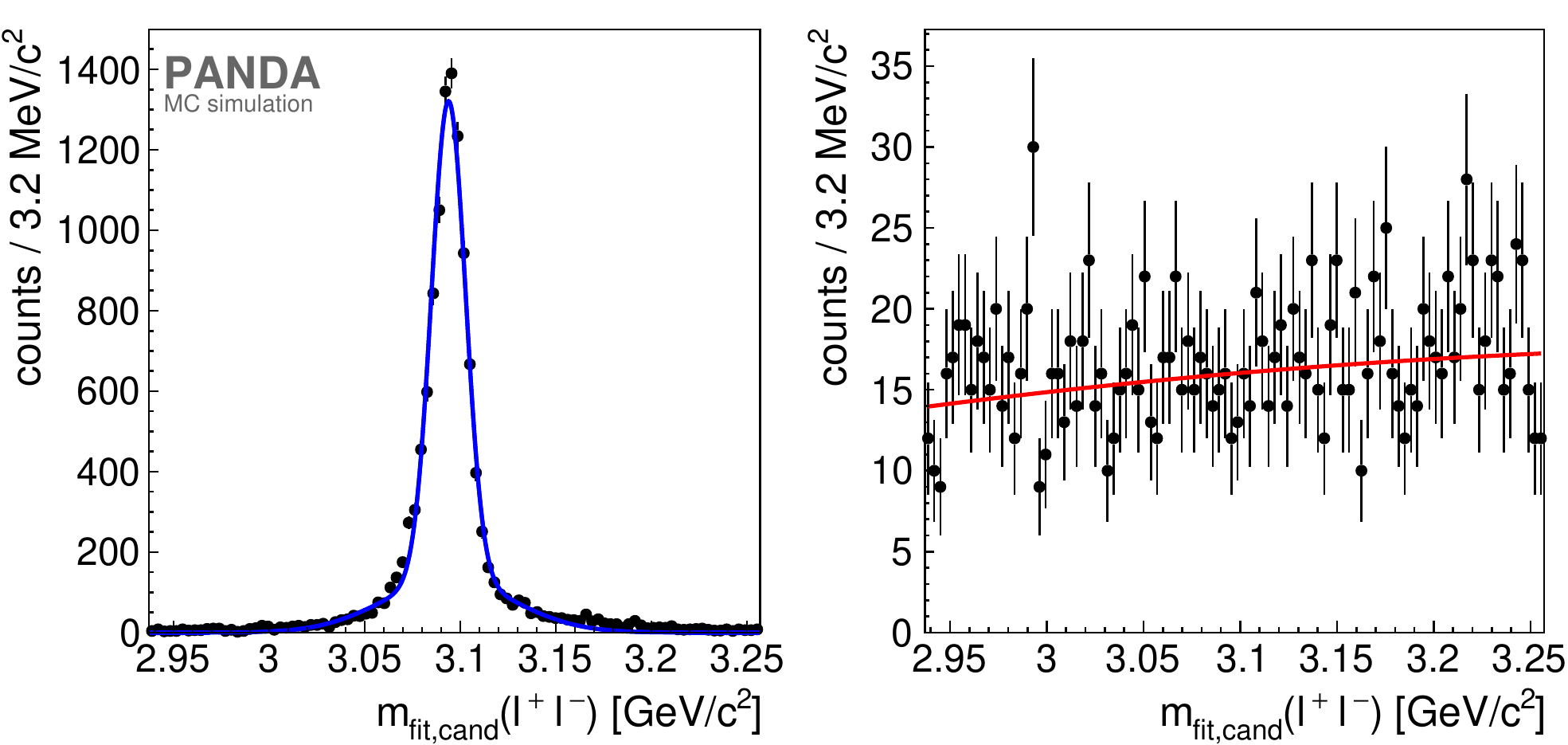}
    \end{center}
      \caption{Determination of the probability density function needed to describe event distributions with $J/\psi$ {\it (left)} 
        and without $J/\psi$ {\it (right)} for the maximum-likelihood approach to determine the energy-dependent event yields. 
        The latter distribution is based on a less tight event selection as compared to the final selection, in order to be able to 
        extract a well defined background shape {\it (right)}.}
\label{fig:MLyields} 
\end{figure}

Furthermore, we need the probability density functions (PDF) of the di-lepton candidate mass distributions for signal and background events. The similarity of the distributions for the two leptonic decay channels allows for a common PDF description for events with ($S$, NR) and without (gen) $J/\psi$ resonance. The PDFs are also extracted from the reconstructed Geant-based MC data. The PDFs for events from signal and non-resonant background (Fig.\,\ref{fig:MLyields}, left) are taken from the di-lepton spectra of the $J/\psi \to e^+e^-$ channel, which due to the slightly worse resolution is more conservative. The generic background PDFs (Fig.\,\ref{fig:MLyields}, right) are taken from $J/\psi \to \mu^+\mu^-$, since the higher remaining number of events allows for a better determination of the background shape.

For each energy scan point $E_{{\rm cms}, i}$, a Poisson random number for the expected event yield is generated, again for each of the three different event types ($S$, NR, gen), respectively, and the di-lepton spectra are generated with these numbers of entries distributed according to the corresponding PDFs. The in this way simulated distributions of the di-lepton spectra are then fitted via an unbinned ML method to extract the corresponding yield of $J/\psi$ events, using {\tt RooFit}~\cite{RooFit}, with the signal and background shapes fixed to those of the PDFs, just leaving the relative event fractions floating.

Finally, the convolved line shape plus constant background level (to account for the non-resonant background) $A\cdot\sigma^{*}(E_{\rm cms},P) + a_{\rm bg}$ is fitted to the resultant energy-dependent event yield distribution to extract the parameter $P$ of interest for the two line shape scenarios under study, namely either $\Gamma_{\rm 0}$ or $E_{\rm f}$. $A$ is the overall amplitude parameter of the fit.

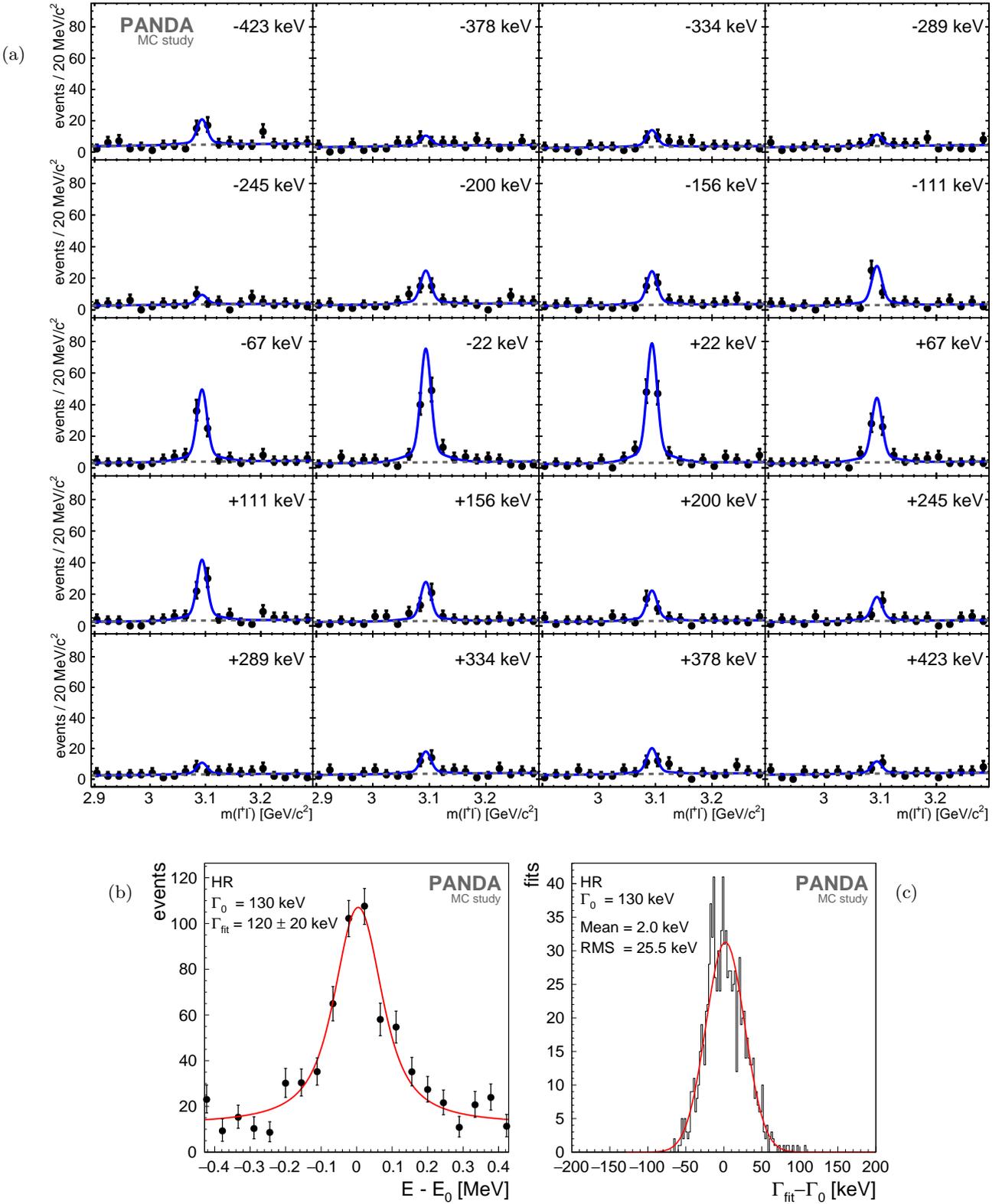
\begin{figure*}[tp]
    \begin{center}
\includegraphics[width=0.9\linewidth,angle=0]{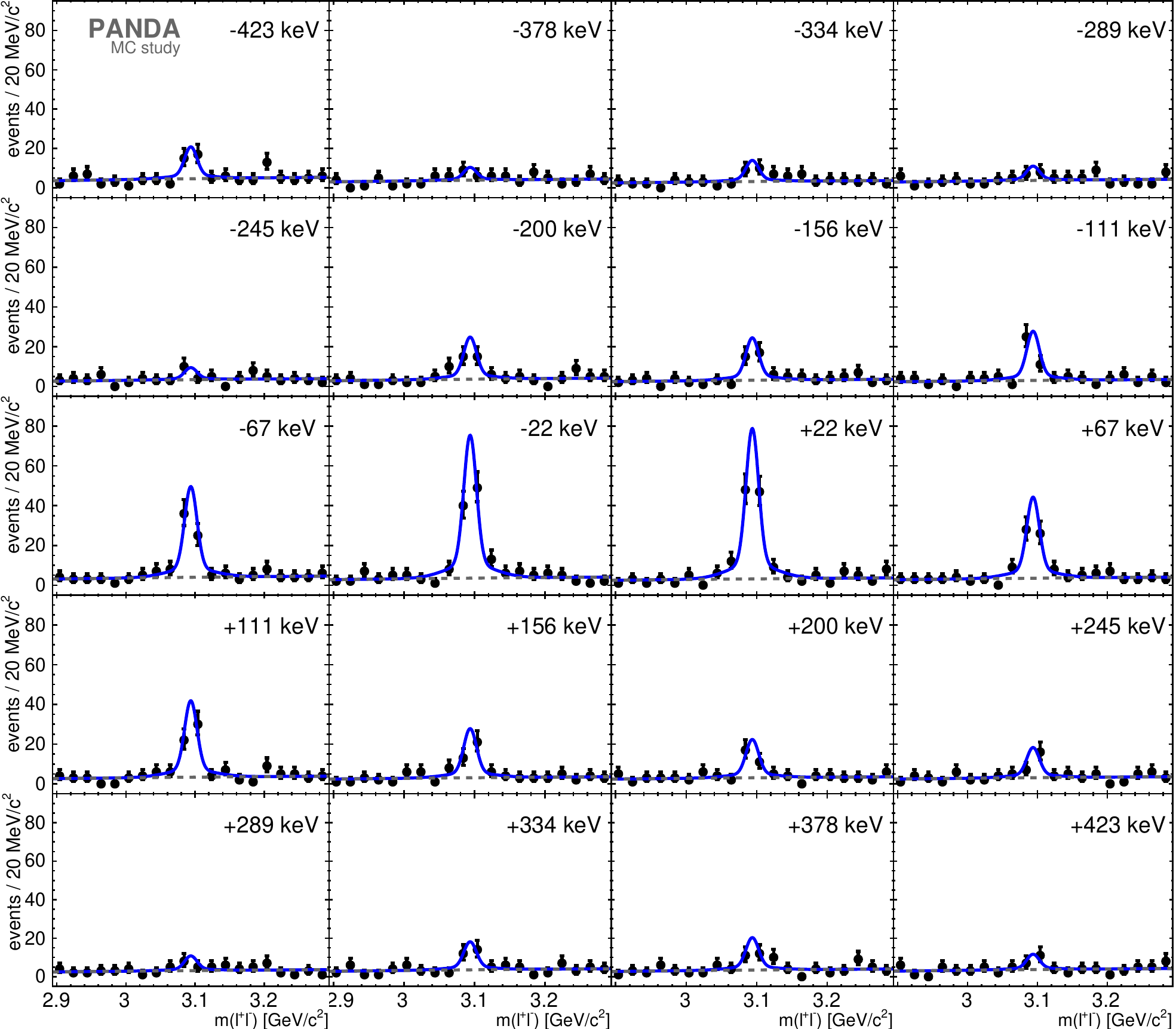}
\vspace{0.3cm}
\end{center}
\begin{center}
\includegraphics[width=0.35\linewidth,angle=0]{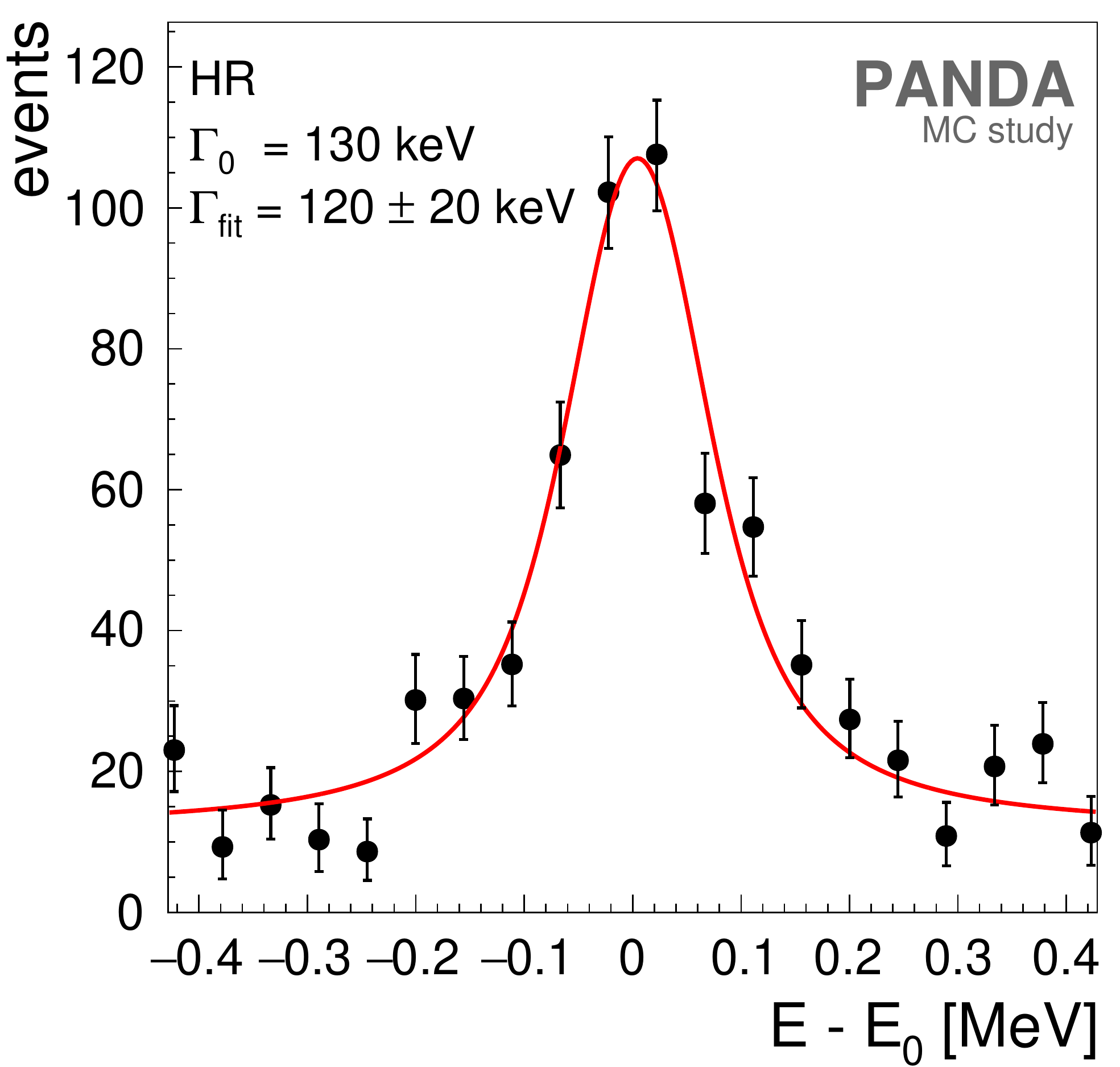}
\includegraphics[width=0.35\linewidth,angle=0]{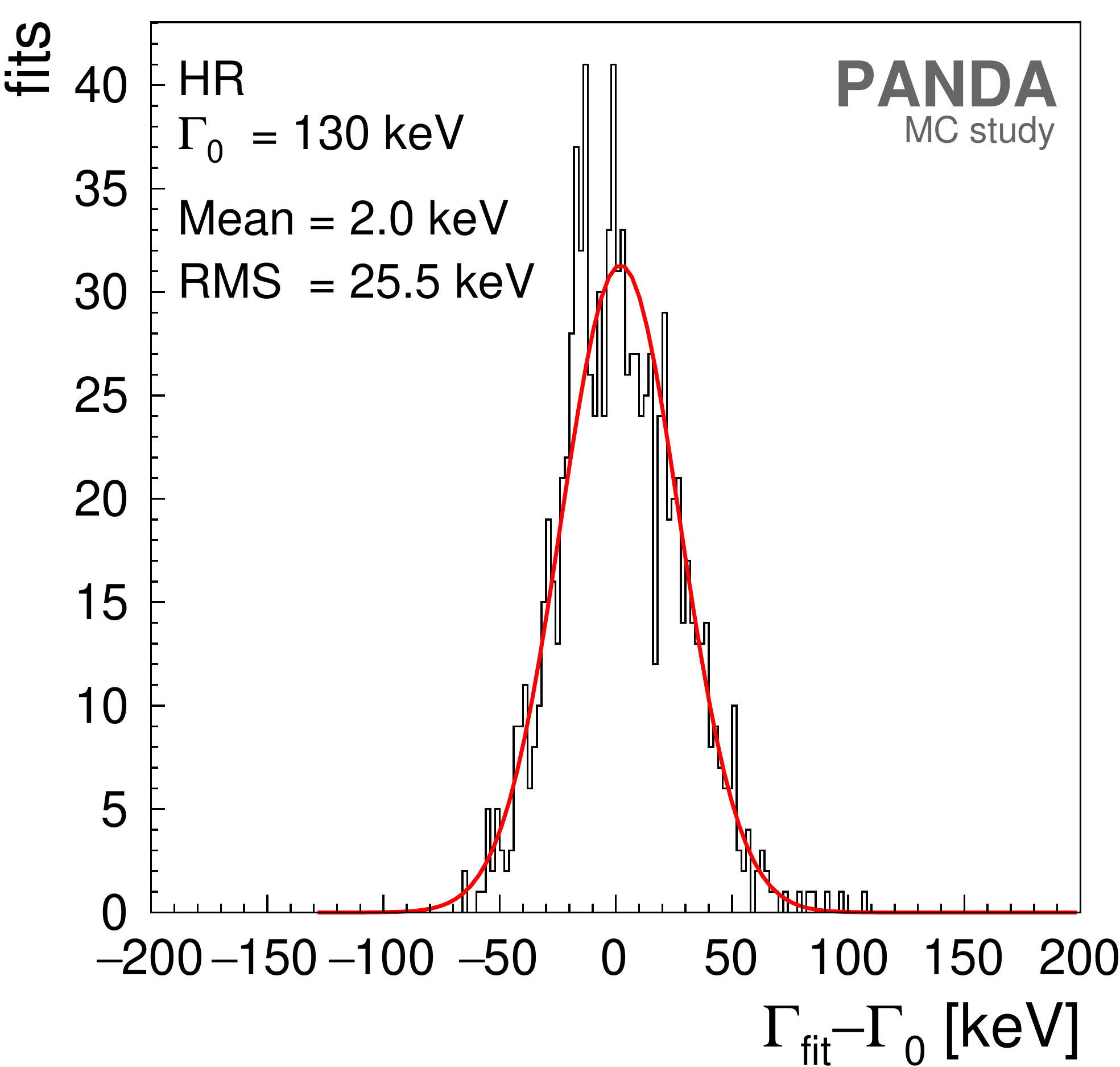}
  \begin{picture}(1,1)
 \put(-437,563){(a)}
 \put(-385,150){(b)}
 \put(0,150){(c)}
  \end{picture}
    \end{center}
      \caption{Illustration of a scan process for the parameter setting: $\sigma_{S} = 100$ nb, $\Gamma_0=130$ keV, 20 energy 
        scan positions (step size $dE\approx 50$ keV), 2 days of data taking per position, HR mode. The set of 20 small plots 
        (a) represent the energy-dependent simulated distributions (going from left to right, top to bottom step-wise through 
        the energy range $(E-E_0)$ shown in (b)) of the reconstructed invariant di-lepton candidate mass containing signal, 
        non-resonant and generic DPM background contributions. (b) shows the resultant energy-dependent yield distribution 
        fitted with a function to extract the parameter of interest, here the Breit-Wigner width $\Gamma$, around the nominal 
        centre-of-mass energy $E_0 = 3.872$ GeV. (c) shows the distribution of this extracted parameter compared to the input 
        value $\Gamma_0$ for 1000 toy MC experiments, allowing the determination of the expected precision (root-mean-square of 
        the distribution) and the accuracy (shift of the distribution). The additional Gaussian function that is fitted to the 
        distribution indicates proper statistical conditions.}
\vspace{-0.3cm}
\label{fig:IlluScan} 
\end{figure*}

Figure~\ref{fig:IlluScan} shows as an example the different steps for an assumed input Breit-Wigner width of $\Gamma_{0}=130\,$keV. The individually generated di-lepton spectra for all $E_{\rm cms}$ (here, for illustration purposes only 20) are shown (Fig.\,\ref{fig:IlluScan}, a), from which the $J/\psi$ yields are extracted to the resultant energy-dependent event yield distribution that is overlaid with the fitted line shape, here a Breit-Wigner function (Fig.\,\ref{fig:IlluScan}, b). 
The distribution of the resultant fit parameter, here the decay width $\Gamma_{0}$, from 1000 performed simulations of this kind with subsequent fit are shown (Fig.\,\ref{fig:IlluScan}, c). The achievable precision is estimated by the root-mean-square (RMS) of the corresponding distribution of results for each case, {\it i.e.} the example input value of $\Gamma_0 = 130$\,keV leads to a measured value of $\Gamma_{\rm meas} = 132.0 \pm 25.5$\,keV. 

The above procedure based on $i=N_{\rm scan}=40$ energy scan points has been simulated using the reconstruction efficiencies (Tab.\,\ref{tab:effi}) for the different HESR operation modes (Tab.\,\ref{tab:Lumis}) for each combination of the various signal input assumptions ($\Gamma_{\rm 0}, \sigma_{S}$) or ($E_{\rm f,0}, \sigma_{S}$) as summarised in Tab.\,\ref{tab:parasum}.

\subsection{Figure of merit for performance studies}
\label{subsec:FoM}
Based on the distributions of the fit results ({\it i.e.} for the measured $\Gamma$ in the Breit-Wigner Case and $E_{\rm f}$ in the Mole{-}cule Case, respectively) from the many MC simulated scan experiments, as exemplarily illustrated in Fig.\,\ref{fig:IlluScan} c), we define our figures-of-merit as follows. 

For the Breit-Wigner Case, the obtained sensitivity is the relative error level $\Delta\Gamma_{\rm meas}/\Gamma_{\rm meas}$ of the measured width, defined as the ratio  of the RMS and the mean value (Mean) shifted by the input width $\Gamma_0$, both taken from the distribution of the fit results $\Gamma_{\rm fit}$\,-$\,\Gamma_{\rm 0}$:
\begin{linenomath*}
\begin{equation}
\label{eq:BW_Sens}
\frac{\Delta\Gamma_{\rm meas}}{{\Gamma}_{\rm meas}} =  \frac{\rm RMS}{{\rm Mean} + \Gamma_0}~~~~~~~~\,{\rm (Breit\mbox{-}Wigner~Case)}~.\\
\end{equation}
\end{linenomath*}

\begin{figure}[tp!]
\centering
  \includegraphics[width=0.34\textwidth]{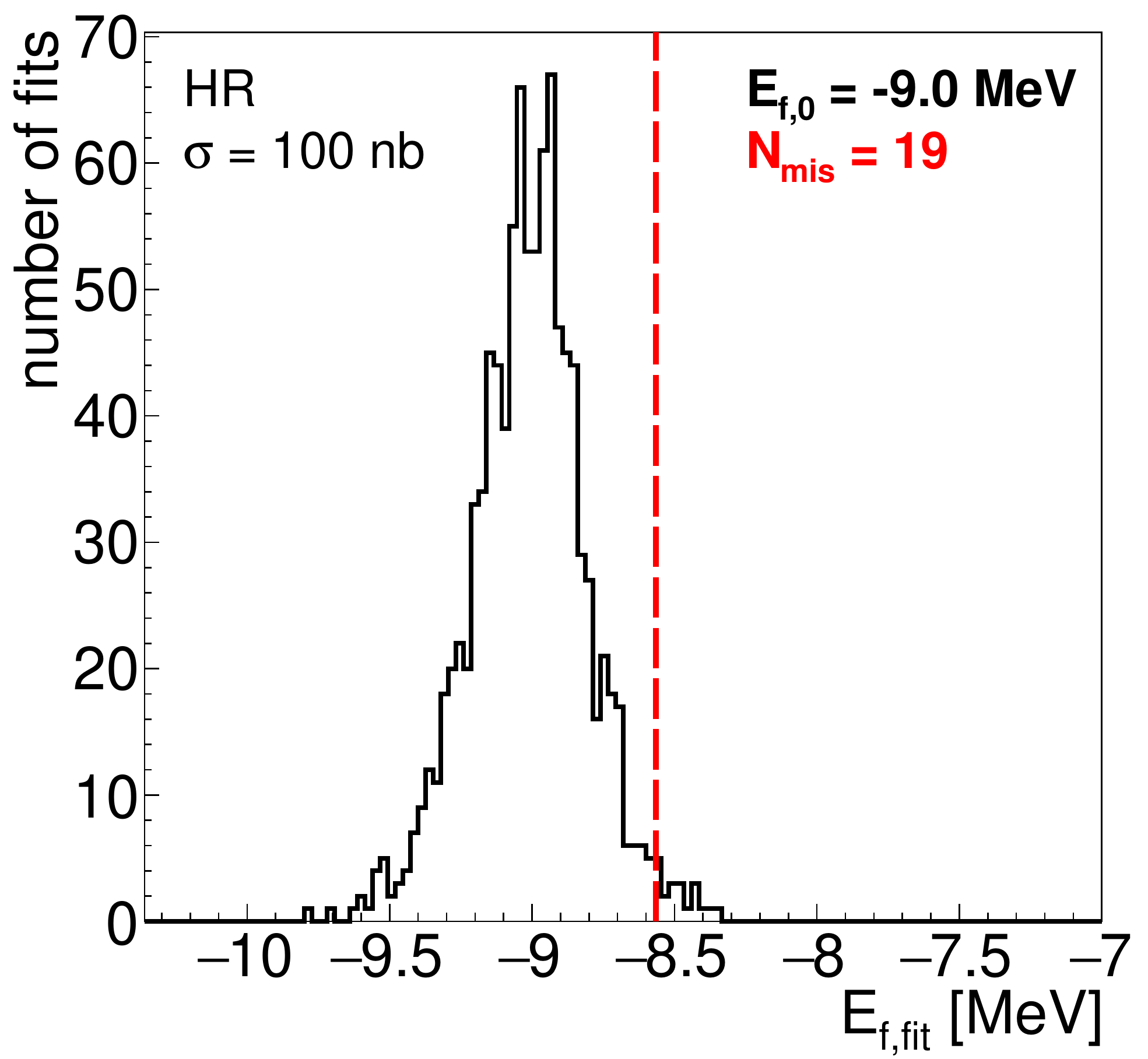}
  \caption{Misidentification fraction for the Molecule Case sensitivity study. Exemplarily shown is the distribution of the fit results 
    for the cross-section assumption $\sigma_{S,{\rm 0}} =$ 100\,nb and the HESR running mode HR for $E_{\rm f,0} = -9.0$\,MeV. The vertical 
    dashed line marks the threshold energy $E_{f,\rm th} \approx -8.5652$\,MeV, where the bound state ($E_{\rm f}<E_{\rm f,th}$) turns into a 
    virtual state. The fraction on the right-hand side of the threshold adds up to $N_{\rm mis} = 19$.}
    \label{fig:MisPID_Fraction_Illus}                                 
\end{figure}

For the Molecule Case, we define the sensitivity as the misidentification probability $P_{\rm mis}$ of the nature of the state. It is determined by computing the fraction of $E_{\rm f}$ fit results being on the ``wrong side'' of the threshold energy $E_{\rm f,th}$, {\it cf.} Sec.\,\ref{subsec:Objectives}:
\begin{linenomath*}
\begin{equation}
\label{eq:Mol_Sens}
P_{\rm mis} = N_{\rm mis}/N_{\rm MC}~~\,{\rm (Molecule~Case)}~,
\end{equation}
\end{linenomath*}
where $N_{\rm MC}$ is the total number of MC simulated scan experiments performed per parameter setting. And $N_{\rm mis}$ is the number of MC experiments with the resulting fitted value $E_{\rm f,meas}$ found on the opposite side of the threshold $E_{\rm f,th}$ than the corresponding input value $E_{\rm f,0}$, leading to misidentification of the true nature of the state, {\it i.e.} confusing the bound with the virtual state scenario and vice versa. This is illustrated in Fig.\,\ref{fig:MisPID_Fraction_Illus}, where the distribution of the fit results is exemplarily shown for a simulated bound state with input parameter $E_{\rm f,0} = -9.0$\,MeV.

The Breit-Wigner sensitivity results are plotted for each HESR running mode (Tab.\,\ref{tab:Lumis}, Sec.\,\ref{subsec:HESR}) {\it vs.} the input parameter $\Gamma_{\rm 0}$ for each signal cross-section $\sigma_{S,{\rm 0}}$ (Tab.~\ref{tab:parasum}, Sec.\,\ref{subsec:ParamSpace}) in Fig.\,\ref{fig:SensitivityResultsBW}. The sensitivity for the distinction between a molecular bound ($E_{\rm f,0}<E_{\rm f,th}$) and a virtual ($E_{\rm f,0}>E_{\rm f,th}$) state as obtained, again, for each HESR running mode (Tab.\,\ref{tab:Lumis}, Sec.\,\ref{subsec:HESR}) {\it vs.} the input Flatt\'{e} parameter $E_{\rm f,0}$ for each signal cross-section $\sigma_{S,{\rm 0}}$ (Tab.~\ref{tab:parasum}, Sec.\,\ref{subsec:ParamSpace}) are summarised in Fig.\,\ref{fig:SensitivityResultsEf}.

\subsection{Treatment of systematic uncertainties}
\label{subsec:Systs}
For estimating the systematic uncertainties, all relevant and, in lack of real data, reasonable sources of uncertainties as listed in Tab.\,\ref{tab:syssources}, partly inspired by a recent comparable BESIII publication~\cite{BesJpsipipi}, are taken into account, adding up to a total uncertainty of about 5\%. They are of different types, and affect the analysis differently. 
%
%
  
\paragraph{Correlated and uncorrelated systematic errors}
\begin{table}[b]
\caption{Sources of systematic uncertainties.}
\vspace{1ex}
\label{tab:syssources}
\centering
\begin{tabular}{c|c|c}
Source & Estimated error $\sigma$ & Total for FS\\\hline
Tracking & 1\%/track  & 4\%  \\
PID & 1\%/identified track & 2\% \\
Kinematic fit & 1\%  & 1\% \\
Bkgd shape & 2\% & 2\% \\
${\cal B}(J/\psi \to l^+l^-)$ & 0.54\% &~~~~~~~0.54\%~\cite{PDG15}\\
Luminosity & 1\% & 1\% \\
\hline
 \multicolumn{1}{c}{}  &  \multicolumn{1}{c|}{Total} & 5.13\% 
\end{tabular}
\end{table}

The uncertainties introduced by tracking and particle identification affect all data sets and types in the same way. They quantify differences between reconstruction efficiencies in ``real'' and MC data, affecting the number of reconstructed events for signal, hadronic background and non-resonant $J/\psi$ background data sets, and are thus correlated, also for all different data sets ``recorded'' at the different $E_{\rm cms}$ scan points. The resultant event yields at all $E_{\rm cms}$ are accordingly scaled by a common factor, while the systematic errors contribute independently to the error of $N_i$ at each energy scan position.

The systematic errors introduced by the 4C kinematic fit, the background shape description as well as the ones on the measured $J/\psi\to\mu^+\mu^-,e^+e^-$ branching fractions (Tab.\,\ref{tab:syssources}) affect only the determined $J/\psi$ yields, and not those of the hadronic background levels. They affect, {\it i.e.} scale, the simulated signal and non-resonant background event yields only, and, again, in the same way for all recorded data sets. Consequently, no systematic error contribution is added to the statistical errors of the individual event yields.

The uncertainty on the luminosity is the only error entering our measurements that affects the data sets at the different centre-of-mass energies independently. Therefore, the event yields obtained at the different energy positions are scaled individually. In consequence, the luminosity error is the only source accounted for in the error bars of the individual event yields.

The resultant systematic uncertainties are determined applying the full analysis procedure in the following way. As for the expected statistical precision (Fig.\,\ref{fig:IlluScan}, c), we repeat once more the scan simulations and cross-section fits 1000 times per parameter setting. Here, the parameters afflicted with uncertainties are randomly varied within the corresponding systematic errors, affecting the simulated expected event yields fit by fit, so that the impact of the systematic uncertainties is taken into account in the subsequently extracted RMS of these fit result distributions. 

\begin{figure*}[tph!]
    \begin{center}
     \includegraphics[clip,trim= 0 0 10 0, width=0.32\linewidth, angle=0]{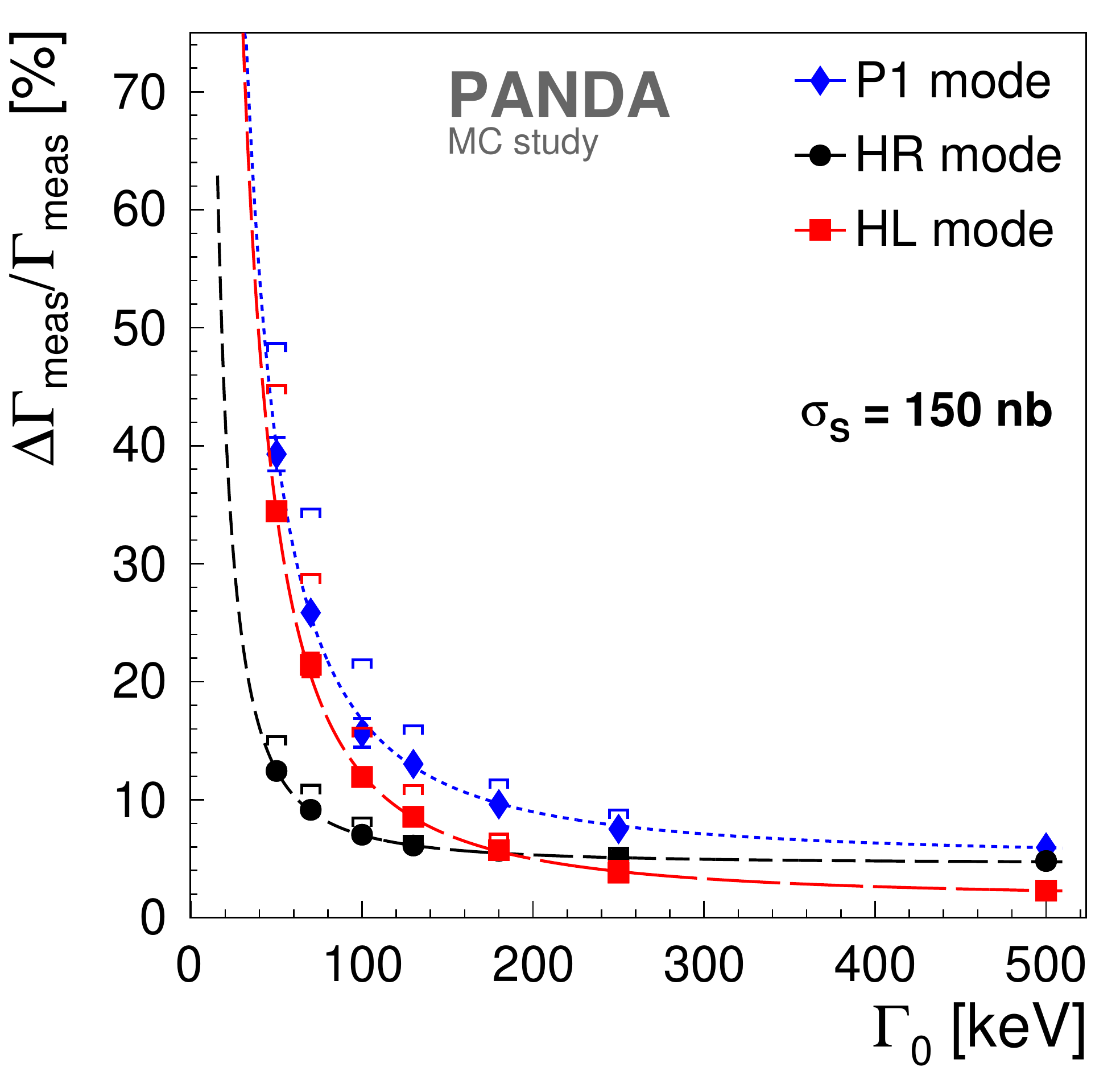}
     \includegraphics[clip,trim= 0 0 10 0, width=0.32\linewidth, angle=0]{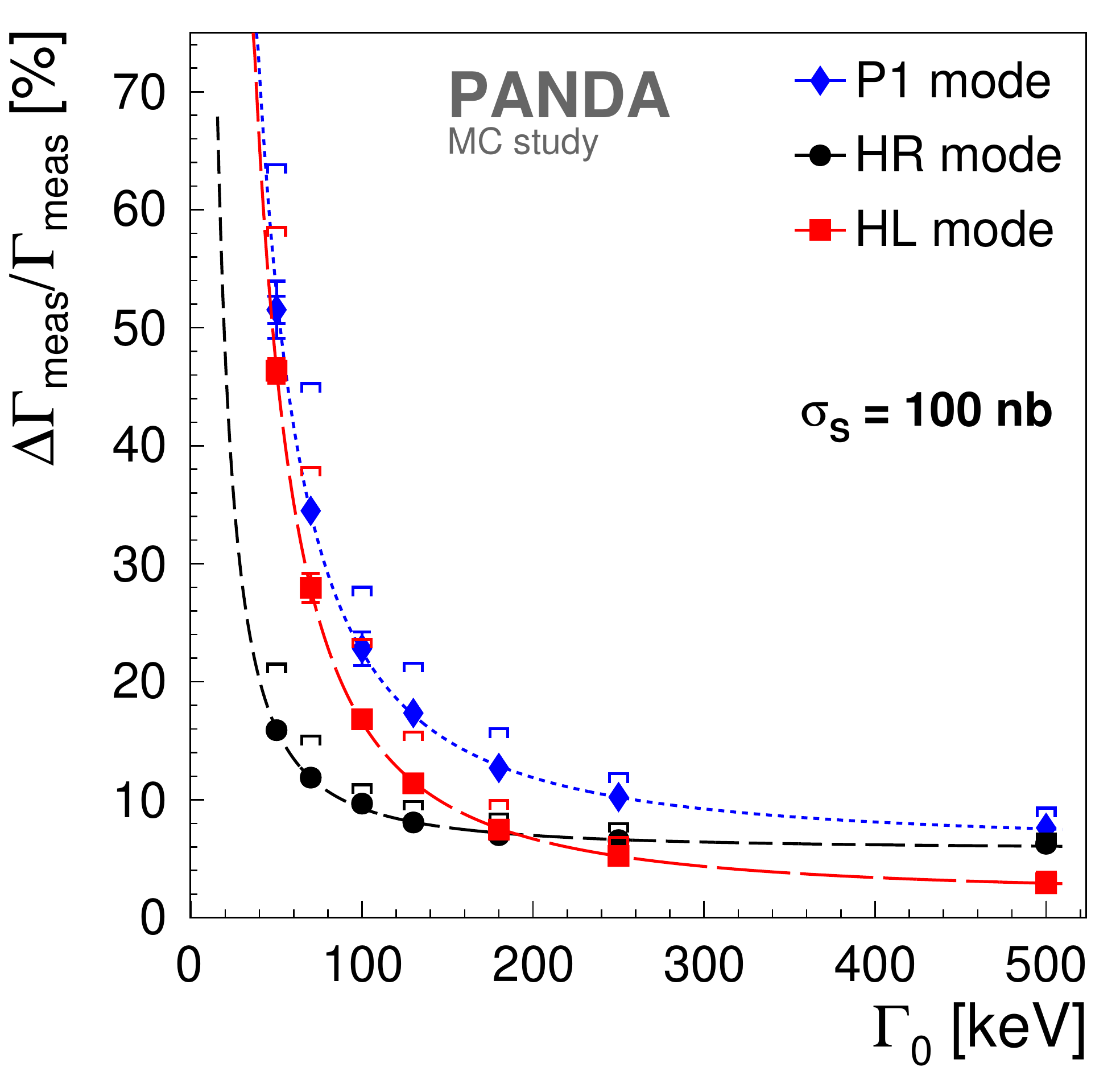}
     \includegraphics[clip,trim= 0 0 10 0, width=0.32\linewidth, angle=0]{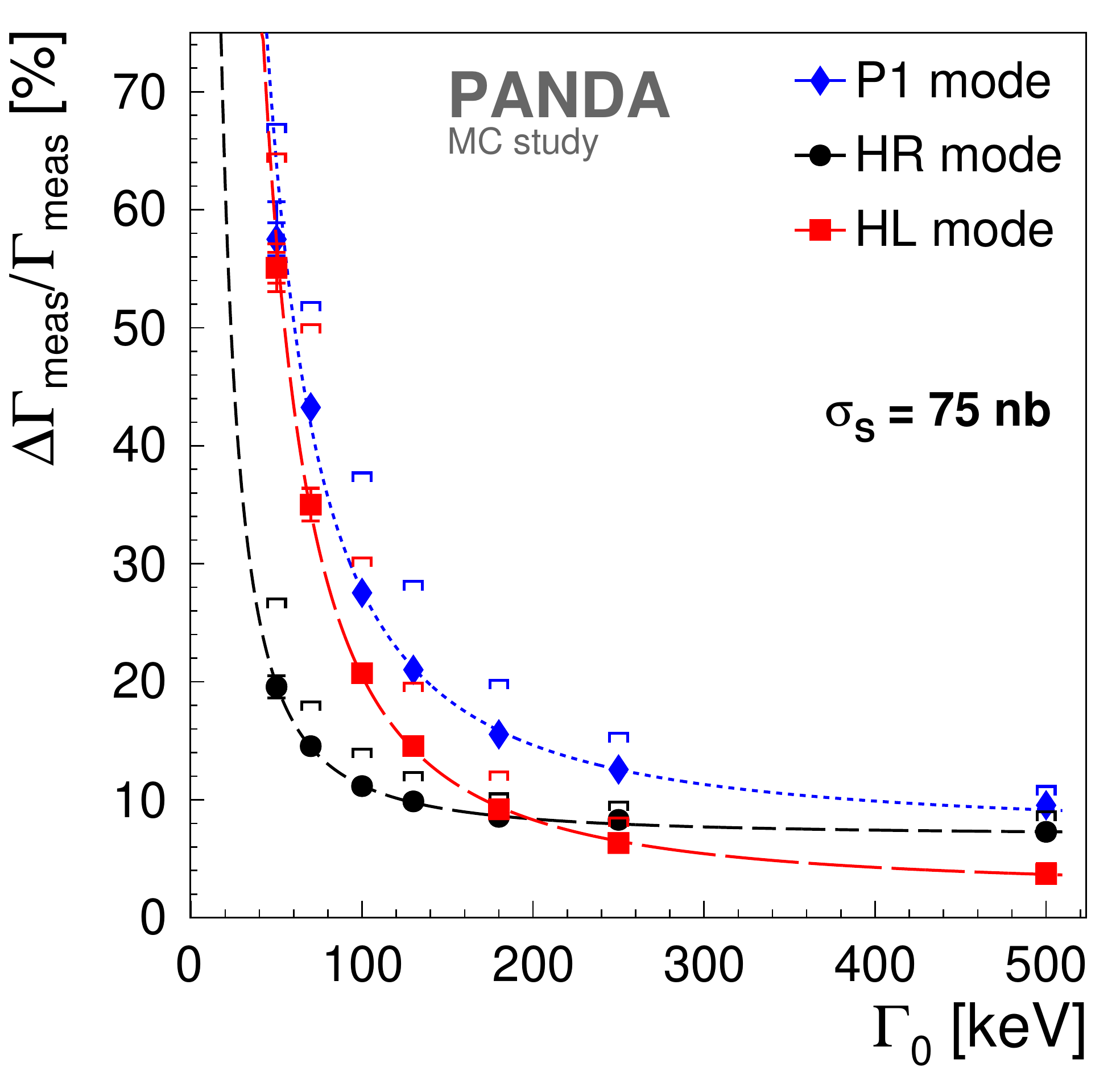}
     \includegraphics[clip,trim= 0 0 10 0, width=0.32\linewidth, angle=0]{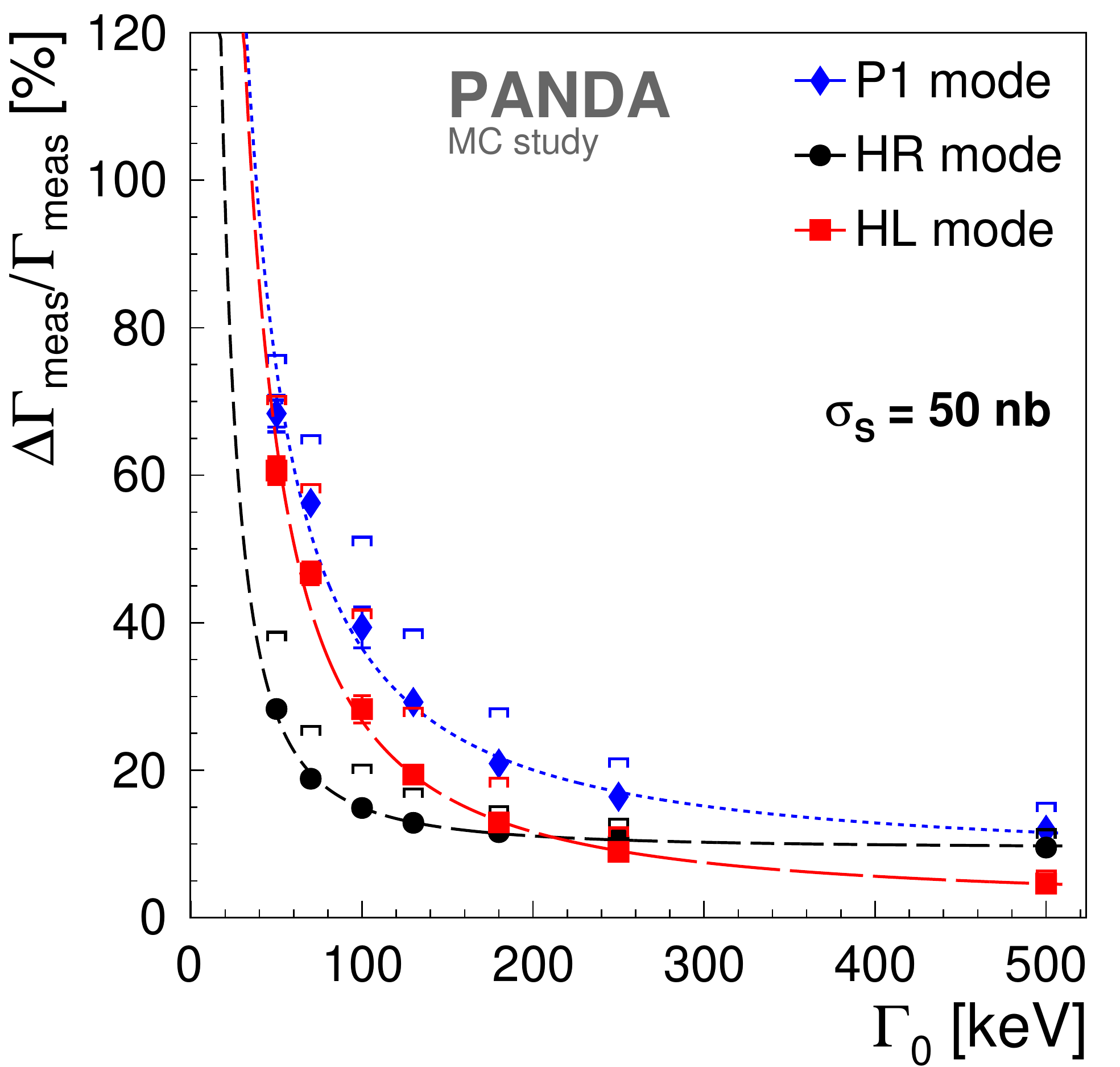}
     \includegraphics[clip,trim= 0 0 10 0, width=0.32\linewidth, angle=0]{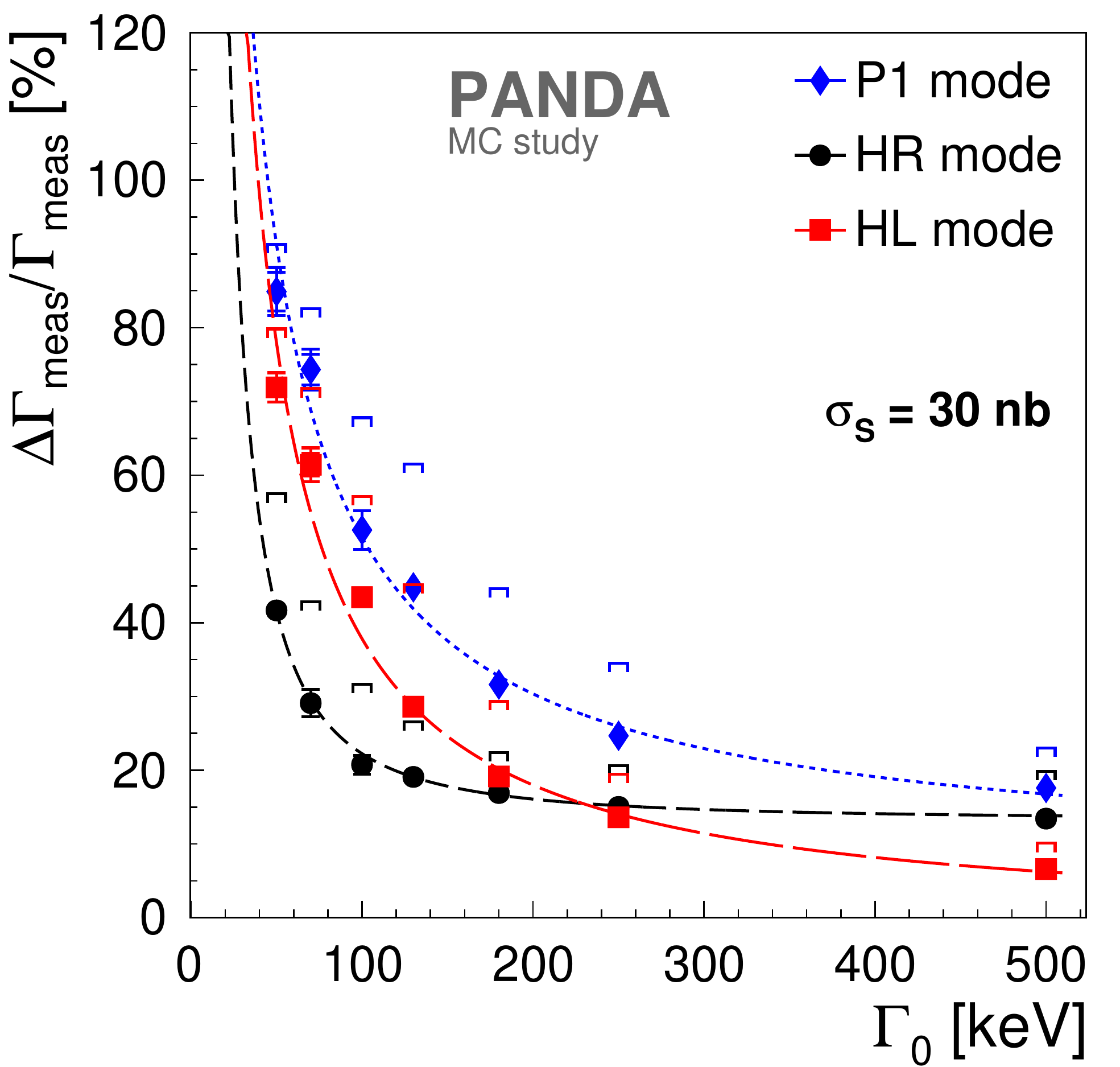}
     \includegraphics[clip,trim= 0 0 10 0, width=0.32\linewidth, angle=0]{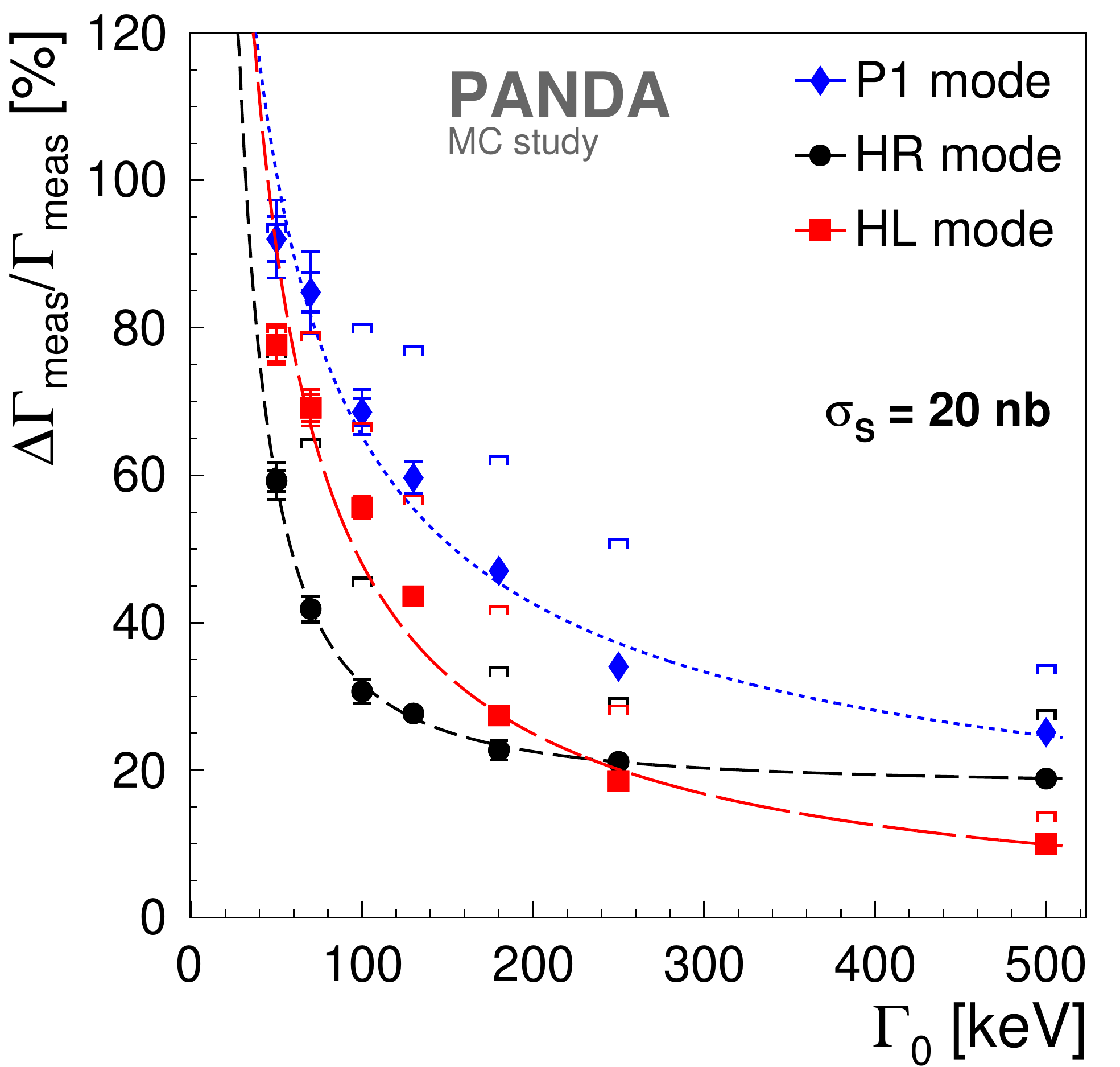}
     \caption{Sensitivity measurements for the Breit-Wigner Case study. Shown are the obtained sensitivities as a function of the assumed 
       natural decay width $\Gamma_{\rm 0}$ of the state under study for the six different signal cross-section assumptions, each for the 
       three different HESR running modes. The inner error bars represent the statistical and the outer ones the systematic errors, and 
       the bracket markers indicate the estimated corresponding numbers for the case of generic hadronic and non-resonant background 
       upscaling (on the data points, ignoring statistical and systematic errors) according to~\cite{OHelene}. 
}
      \label{fig:SensitivityResultsBW}                                 
     \end{center}
\end{figure*}

\begin{figure*}[tp!]
    \begin{center}
     \vspace{-0.2cm}
     \includegraphics[clip,trim= 0 0 10 0, width=0.32\linewidth, angle=0]{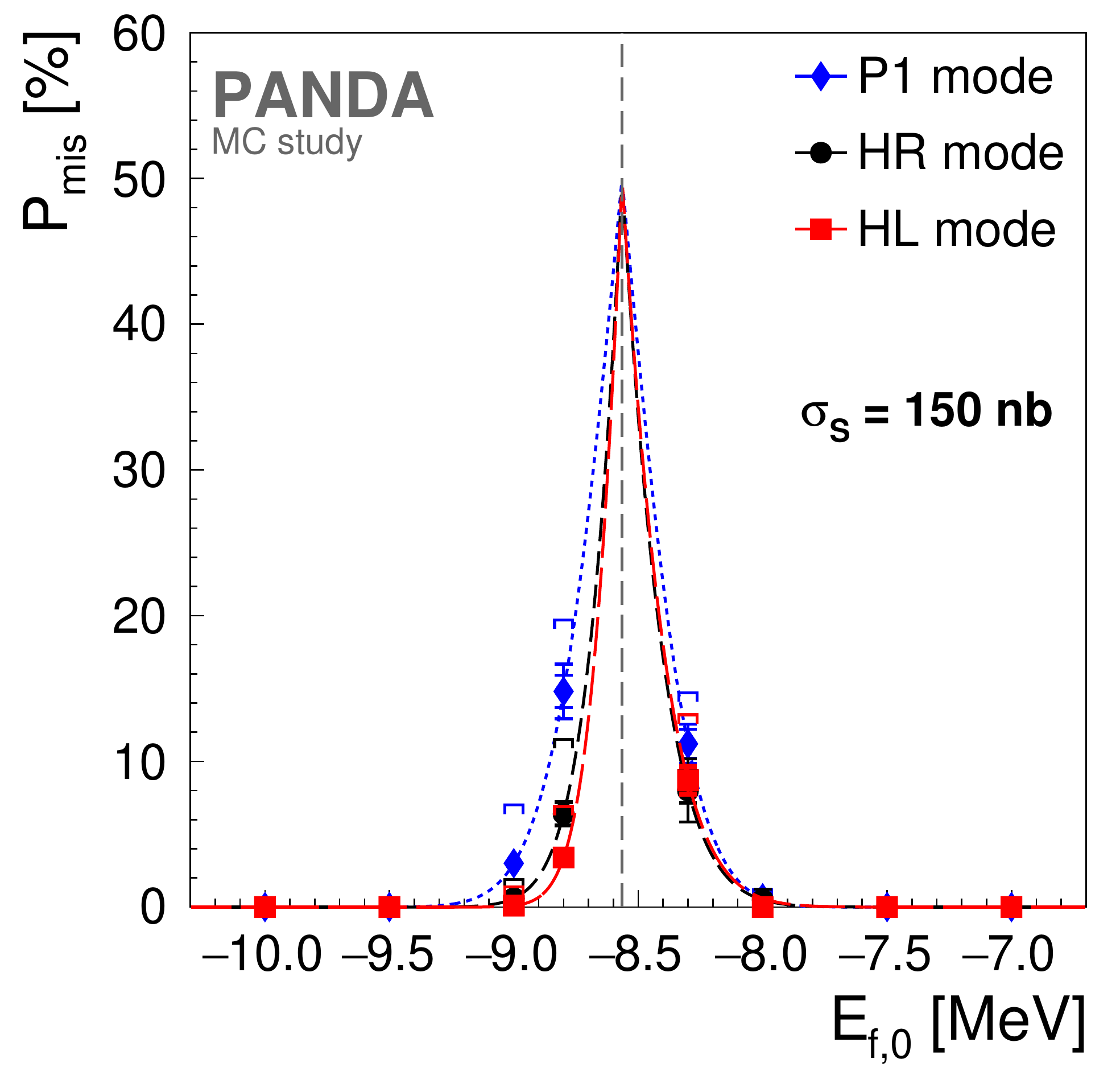}
     \includegraphics[clip,trim= 0 0 10 0, width=0.32\linewidth, angle=0]{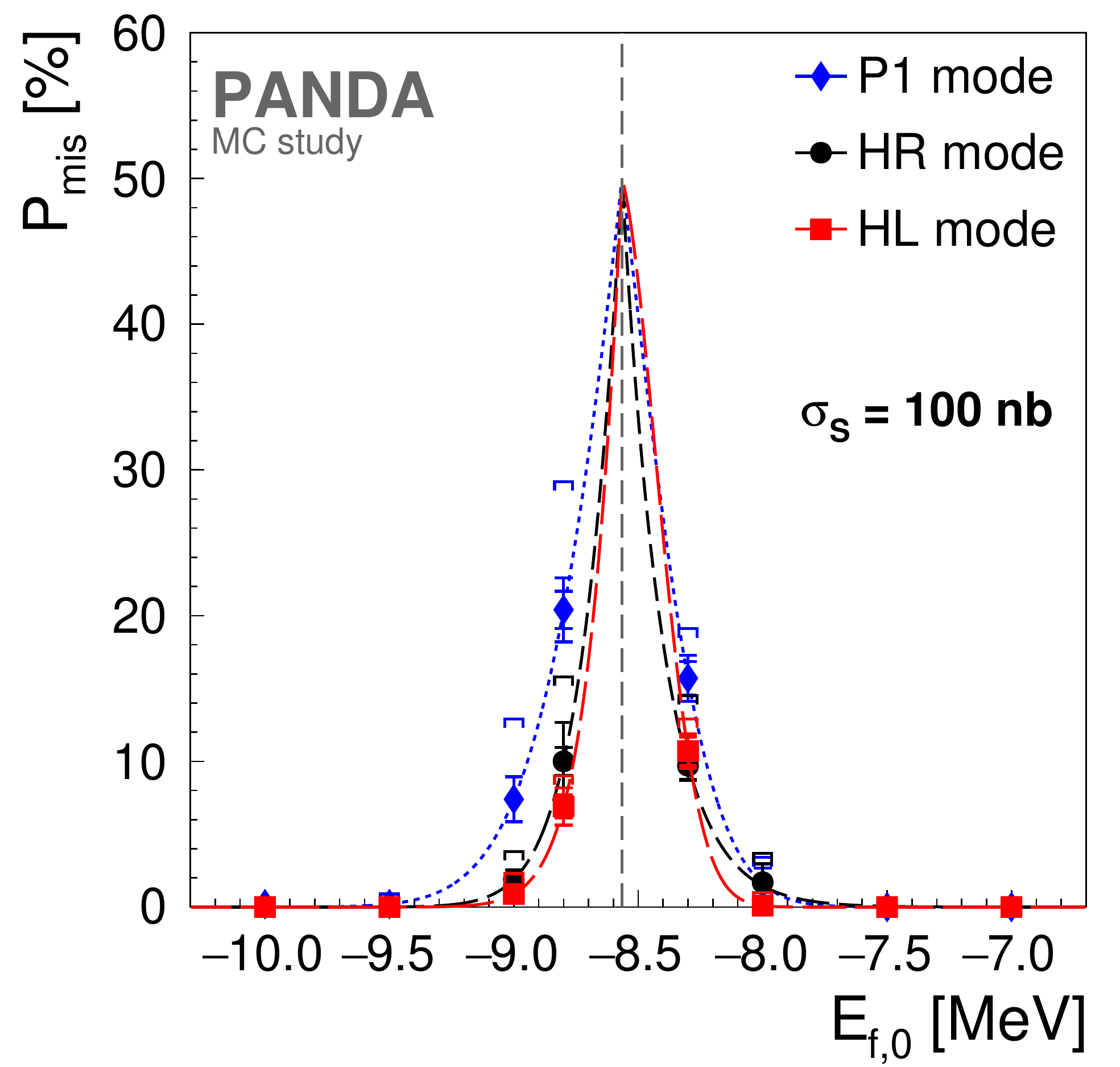}
     \includegraphics[clip,trim= 0 0 10 0, width=0.32\linewidth, angle=0]{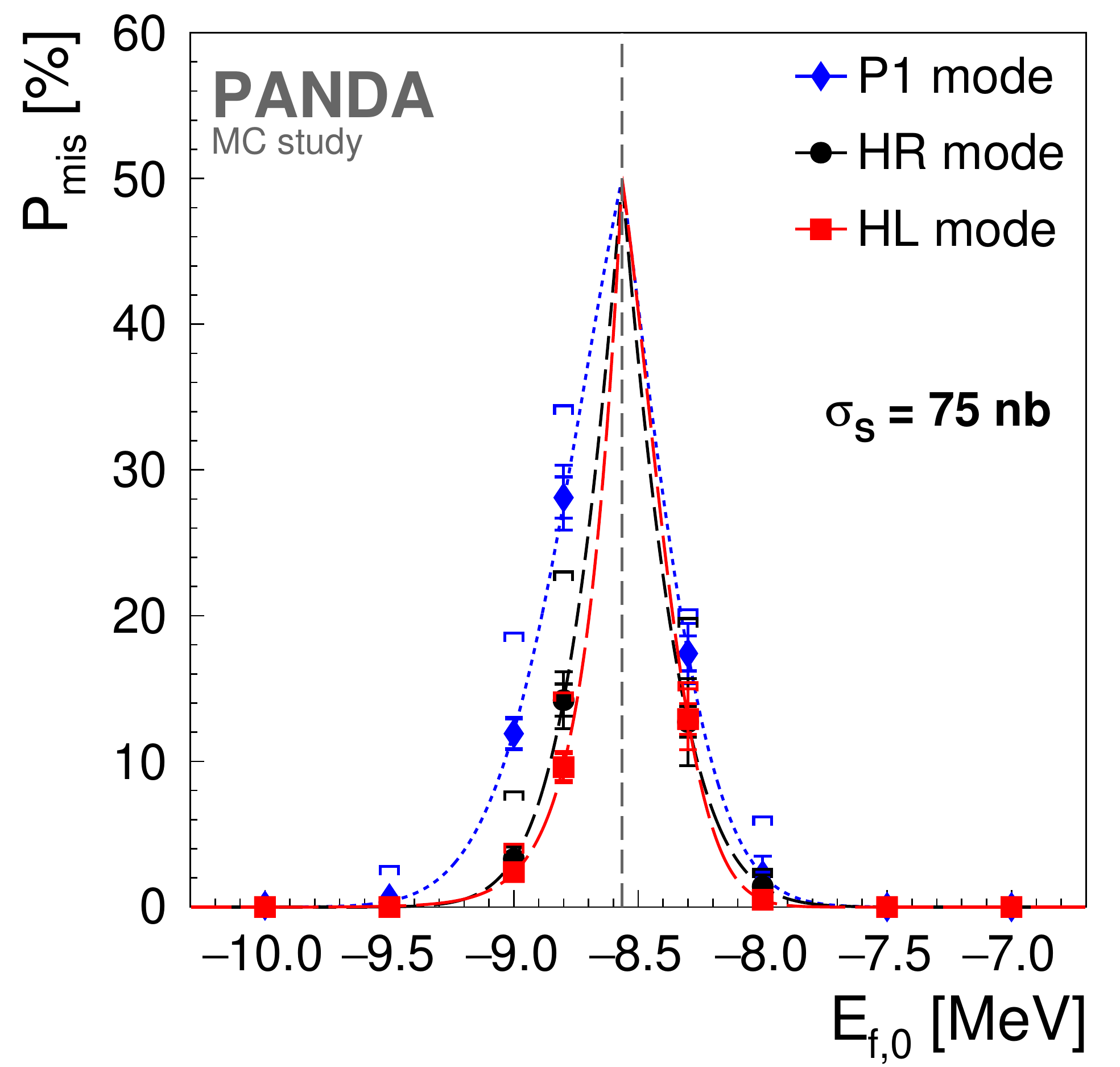}
     \includegraphics[clip,trim= 0 0 10 0, width=0.32\linewidth, angle=0]{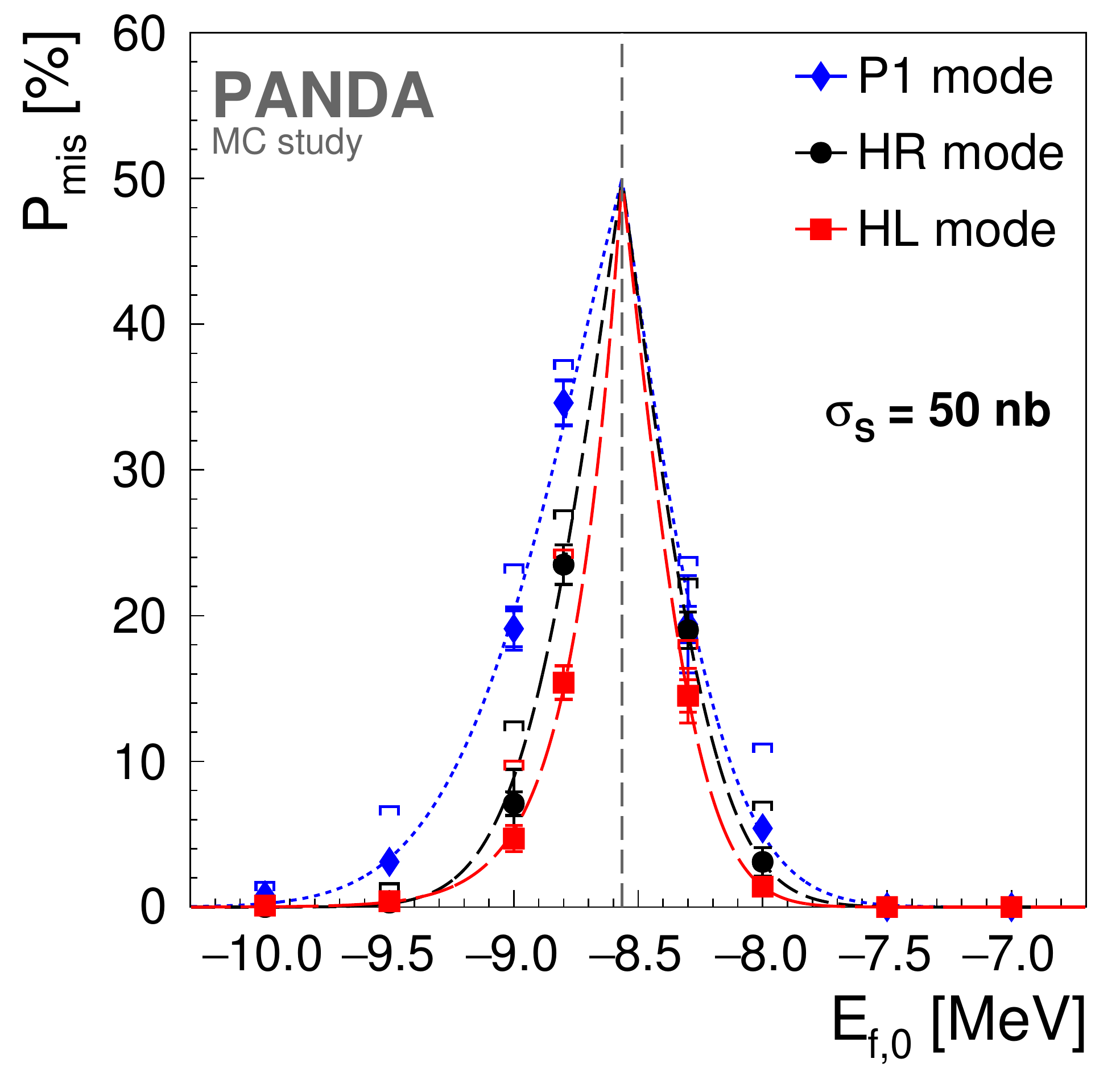}
     \includegraphics[clip,trim= 0 0 10 0, width=0.32\linewidth, angle=0]{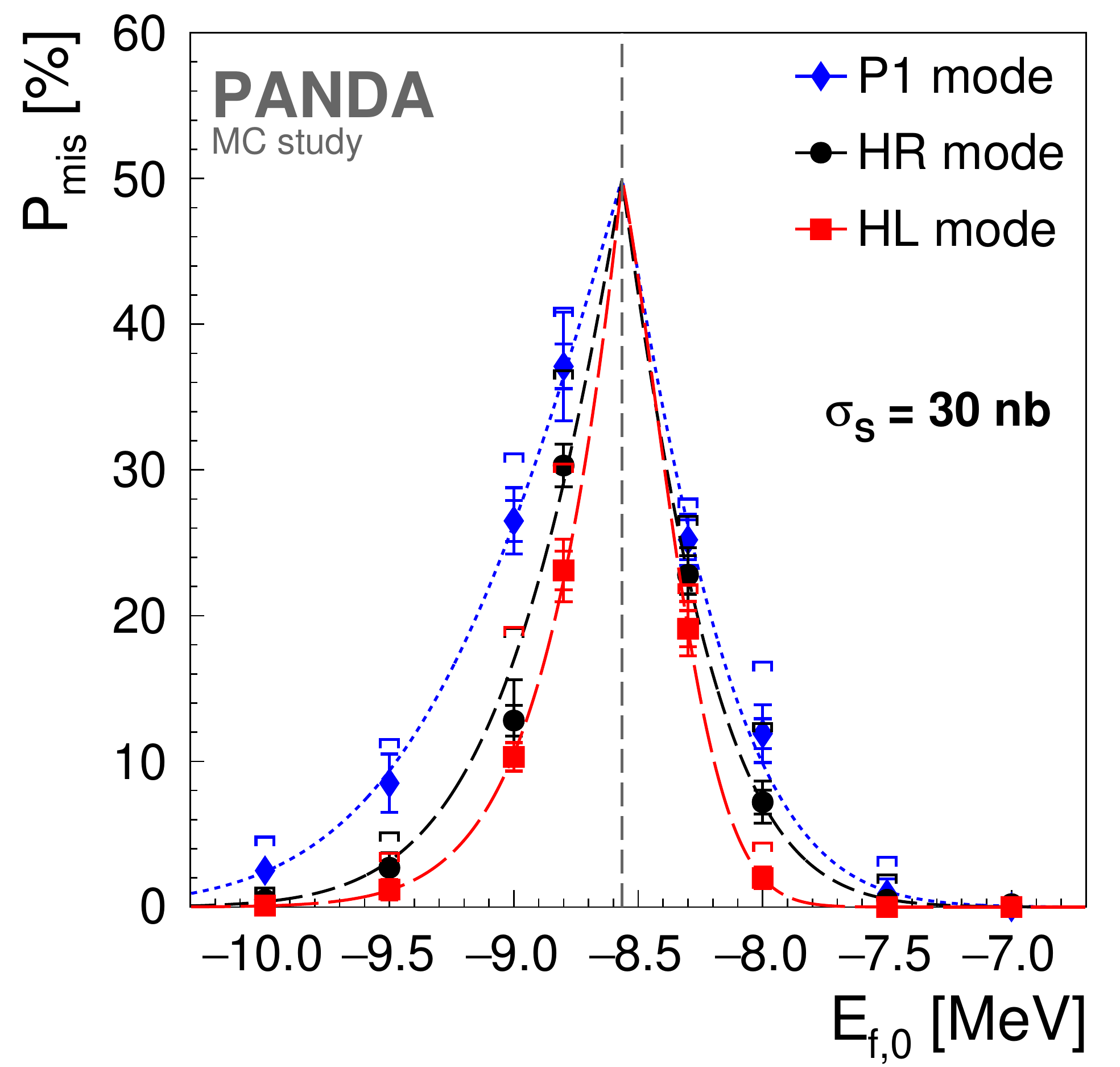}
     \includegraphics[clip,trim= 0 0 10 0, width=0.32\linewidth, angle=0]{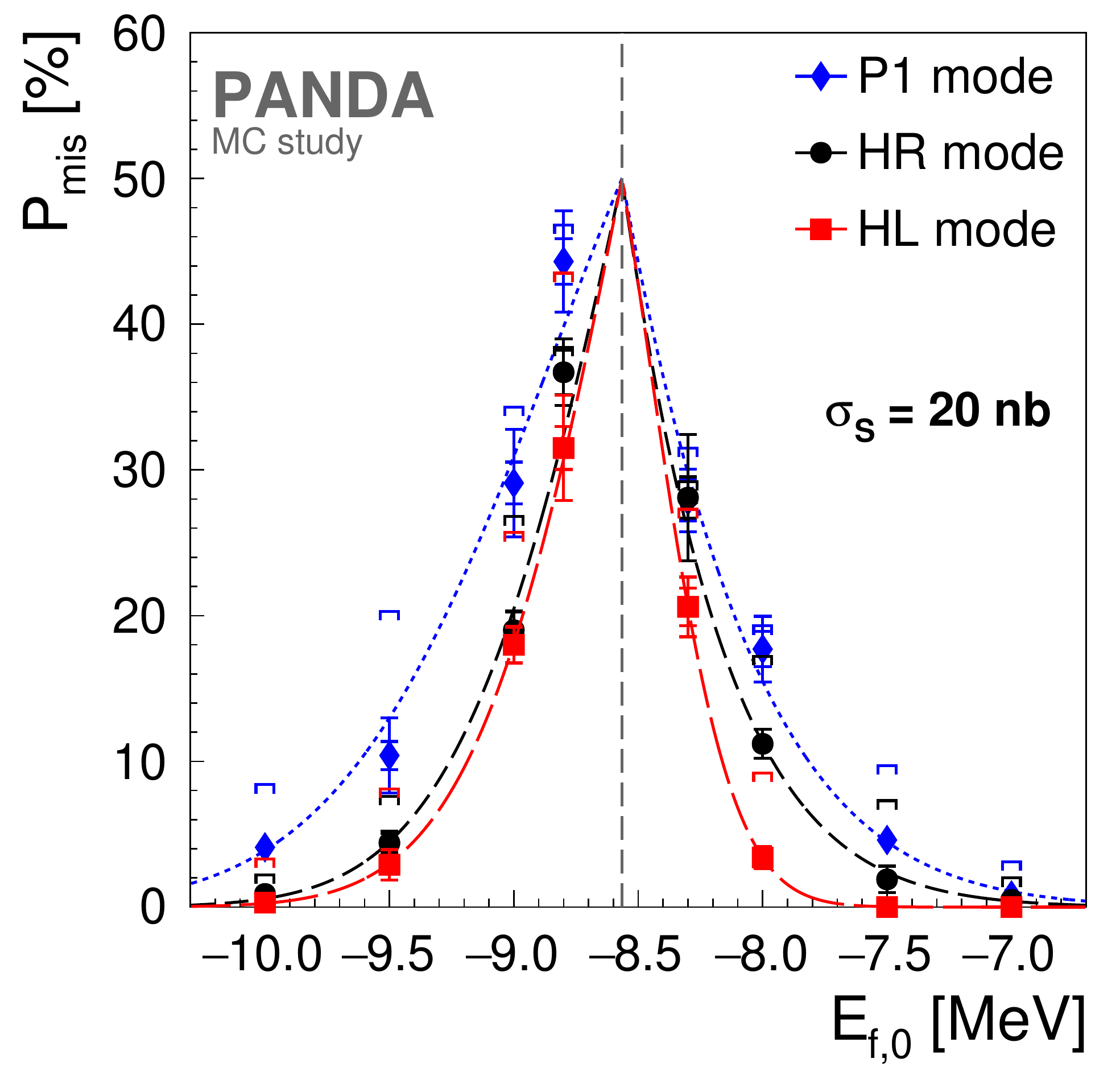}
     \caption{Sensitivity measurements for the Molecule Case study. Shown are the obtained sensitivities in terms of the 
       misidentification probability $P_{\rm mis}$ as a function of the assumed 
       Flatt\'{e} parameter $E_{\rm f,0}$ of the state under study for the six different signal cross-section assumptions, each for the 
       three different HESR running modes. The inner error bars represent the statistical and the outer ones the systematic errors, and 
       the bracket markers indicate the estimated corresponding numbers for the case of generic hadronic and non-resonant background 
       upscaling (on the data points, ignoring statistical and systematic errors) according to~\cite{OHelene}.}
      \label{fig:SensitivityResultsEf}                                 
     \end{center}
\end{figure*}

We take the absolute difference $|\Gamma_{\rm meas}-\Gamma_{\rm meas, incl. syst.}|$ or $|E_{\rm f,meas}-E_{\rm f,meas, incl. syst.}|$ as systematic error estimate and add it in quadrature to the statistical one. In the summarising sensitivity plots (Figs.\,\ref{fig:SensitivityResultsBW} and \ref{fig:SensitivityResultsEf}), the systematic uncertainties are given by the outer error bars. 

In summary, we conclude that the effect of systematic uncertainties is not significant. This is expected because most of the systematic error sources lead to common scaling factors for all scanned event yields. This does not change the signal line shape, and affects the statistical uncertainties only insignificantly. We estimate this effect of a common scaling of all yields by about 5\% (assuming all systematic errors given in Tab.\,\ref{tab:syssources} are uncorrelated) to be a relative change of only about 2.5\% ($\sqrt{1.05 \cdot N}/\sqrt{N} = \sqrt{1.05} \approx 1.025$) in the statistical error of each yield. 
We conclude that this contribution is thus negligible. 
%
%
%
%

\paragraph{Uncertainty due to generic hadronic background}
There is an uncertainty in our sensitivity study results due to the fact that our generic hadronic background MC samples (type ``gen'' in Tab.\,\ref{tab:data}) are (CPU-wise) limited, {\it cf.} Sec.\,\ref{subsec:MCsim}. In consequence, the residual number of generic background events after event selection is quite small 
\linebreak 
($N_{B,ee} = 1$, $N_{B,\mu\mu} = 4$) due to the high background suppression of order $\epsilon_{B,\rm gen}\approx 1\cdot10^{-10}$ (Tab.\,\ref{tab:effi}), introducing uncertainties concerning the generic background levels estimated here {\it vs.} the anticipated future analysis of real data. 

In order to investigate the impact, the number of gen{-}eric background events is scaled up by a factor determined from the 90\% confidence level upper limit (UL90) for the actual count numbers, according to~\cite{OHelene}:
\begin{eqnarray*}
N_{B,ee} = 1  \Rightarrow  {\rm UL90}(N_{B,ee}) & \approx &  4 
  \Rightarrow \epsilon^{\rm up}_{B,ee} = 4\cdot \epsilon_{B,ee}\,,\\
N_{B,\mu\mu} = 4 \Rightarrow  {\rm UL90}(N_{B,\mu\mu}) & \approx &  8  
	\Rightarrow \epsilon^{\rm up}_{B,\mu\mu} = 2\cdot \epsilon_{B,\mu\mu}\,.
\end{eqnarray*}
Applying these factors on the background reconstruction efficiencies and re-performing the 1000 MC scan experiments and fits, a conservative number for the figure of merit is estimated. Practically, the determined $J/\psi$ yield at each energy scan point is affected by the increased background $B_{\rm gen}$, the efficiency-weighted mean value of the scaling factors is $\epsilon^{up}_{B,{\rm gen}} = (4 \cdot \epsilon_{B,ee} + 2 \cdot \epsilon_{B,\mu\mu}) / (\epsilon_{B,ee} + \epsilon_{B,\mu\mu}) \approx 2.4$, {\it cf.} Tab.\,\ref{tab:effi}. The uncertainty of the DPM model itself is difficult to estimate. As indicated by a comparison between DPM~\cite{DPM} predictions and HERA cross-section measurements~\cite{PbarXCERN}, showing for typical background channels of {\it e.g.} $\bar{p}p \to 2\pi^+2\pi^-$ or $\bar{p}p \to 2\pi^+2\pi^-\pi^0$ that DPM overestimates the experimental data by a factor 2-3, we assume it here to be well covered by the error estimate for the upscaling. In this respect, the resultant sensitivities for the resonance energy scan measurements are rather conservative.

\paragraph{Uncertainty due to non-resonant background}
Another limitation of our study is the assumption that the di-pion system of the non-resonant background $\bar{p}p \rightarrow J/\psi\pi^+\pi^-$ (type ``NR'' in Tab.\,\ref{tab:data}) is never produced via an intermediate $\rho^0(770)$. Therefore, we performed the study again, assuming this happens in 50\% of the cases, leading to background events being kinematically indistinguishable from signal events. To account for the impact, the reconstruction efficiency for the non-resonant background $\epsilon_{B,{\rm NR}}$ is scaled up to the average of the original value (without $\rho^0(770)$) and the one of the signal reconstruction efficiency $\epsilon_{S}$ (with $\rho^0(770)$), separately for the electronic and the muonic $J/\psi$ decays to $(2.8\% + 12.2\%)/2 = 7.5\%$ and $(3.0\% + 15.2\%)/2 = 9.1\%$, respectively.  

The impact due to the effectively higher $J/\psi$ yield disconnected from the resonant $X(3872)$ production, {\it i.e.} the higher background level in the scan graph (Fig.\,\ref{fig:IlluScan}, a), is estimated. For this purpose, an effective average scaling factor of $(7.5/2.8 + 9.1/3.0)/2 \approx 2.9$ is applied to $\epsilon_{B,{\rm NR}}$. The upscaling is done at the same time as the upscaling adressing the limited generic hadronic background described previously. 

Applying the factors and re-performing 1000 fits under these modified conditions lead to more conservative results for the extracted sensitivities. The bracket markers in the summarising sensitivity plots (Figs.\,\ref{fig:SensitivityResultsBW} and \ref{fig:SensitivityResultsEf}) 
indicate the corresponding numbers due to the upscaling of the two background contributions according to~\cite{OHelene}.   

\paragraph{Trendlines in sensitivity result plots}
In order to roughly guide the eye for inter- and extrapolations between the actual ``measured'' data points (Figs.\,\ref{fig:SensitivityResultsBW} and \ref{fig:SensitivityResultsEf}), empirical trendlines are fitted to the points (with errors). The extrapolation to smaller input $\Gamma_0$ (Fig.\,\ref{fig:SensitivityResultsBW}), however, are lacking any serious basis and should be taken with a ``grain of salt''. For the trendlines of $P_{\rm mis}$ {\it vs.} $E_{\rm f,0}$ (Fig.\,\ref{fig:SensitivityResultsEf}), the empirical fit functions have been constrained to $P_{\rm mis}=50$\% at $E_{\rm f,0}=E_{\rm f,th}$ according to the statistical definition. At $E_{\rm f,th}$, the model itself does not distinguish between virtual and bound states.  
%
%
\begin{figure}[tp!]
\centering
  \includegraphics[width=0.35\textwidth]{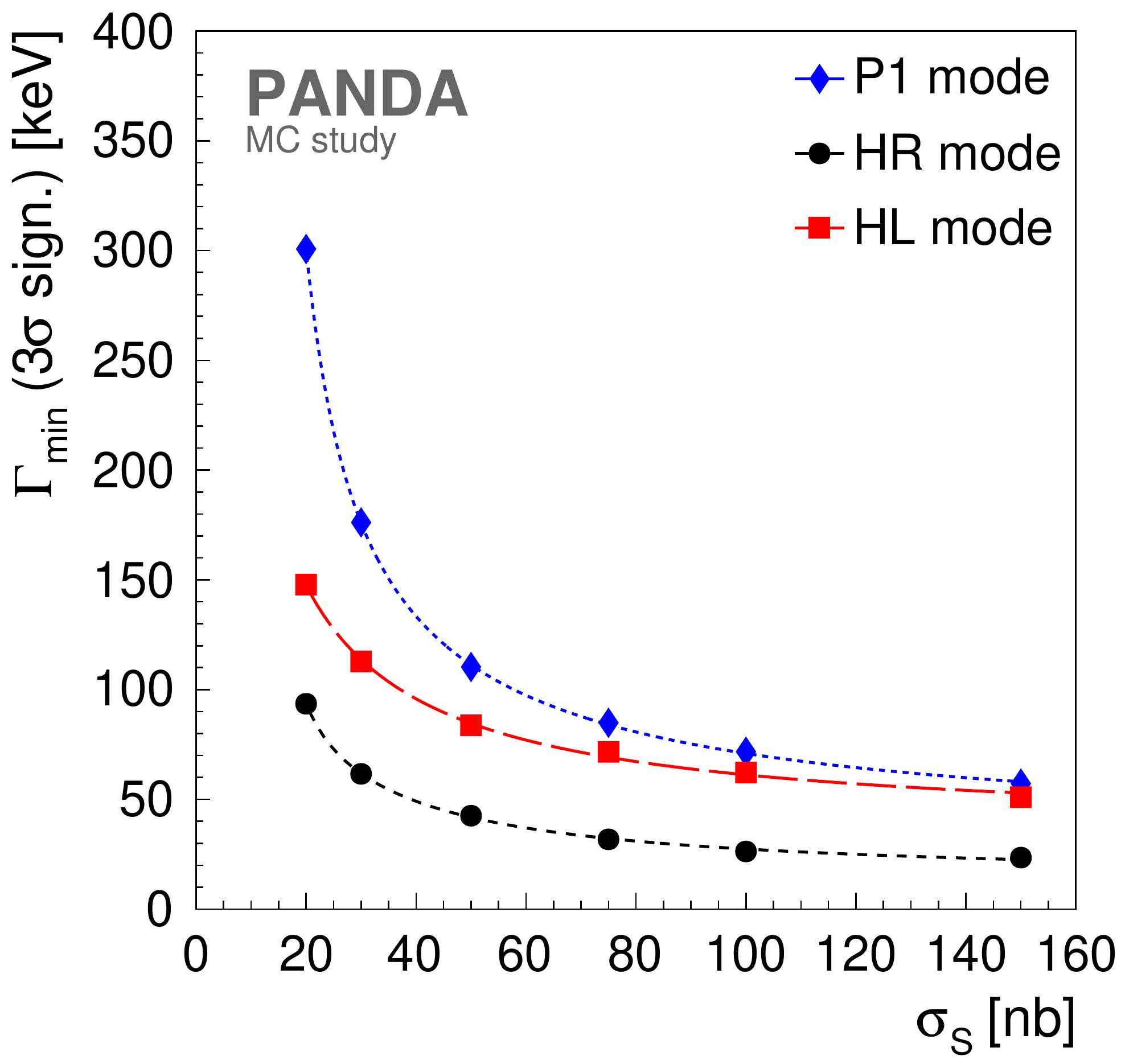}
  \caption{Sensitivity plot for a 33\% relative error ($3\sigma$) Breit-Wigner width measurement.}
  \label{fig:scanResult_3sigma_BW_trend}
\end{figure}

\subsection{Discussion of results}
\label{subsec:Results}
%
%
\begin{figure*}[tp!]
\centering
  \includegraphics[width=0.35\textwidth]{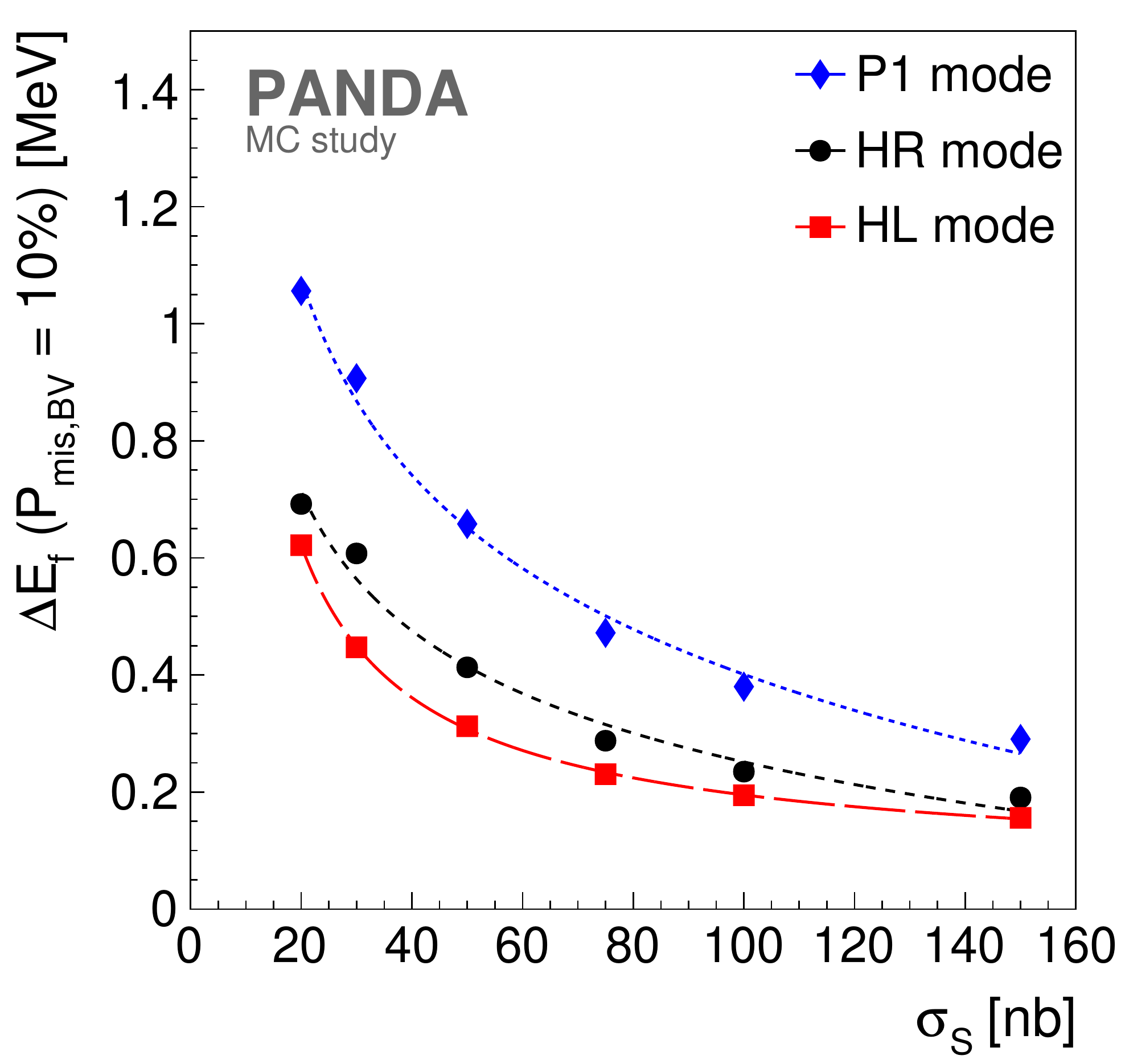} 
\hspace{1.2cm}
  \includegraphics[width=0.35\textwidth]{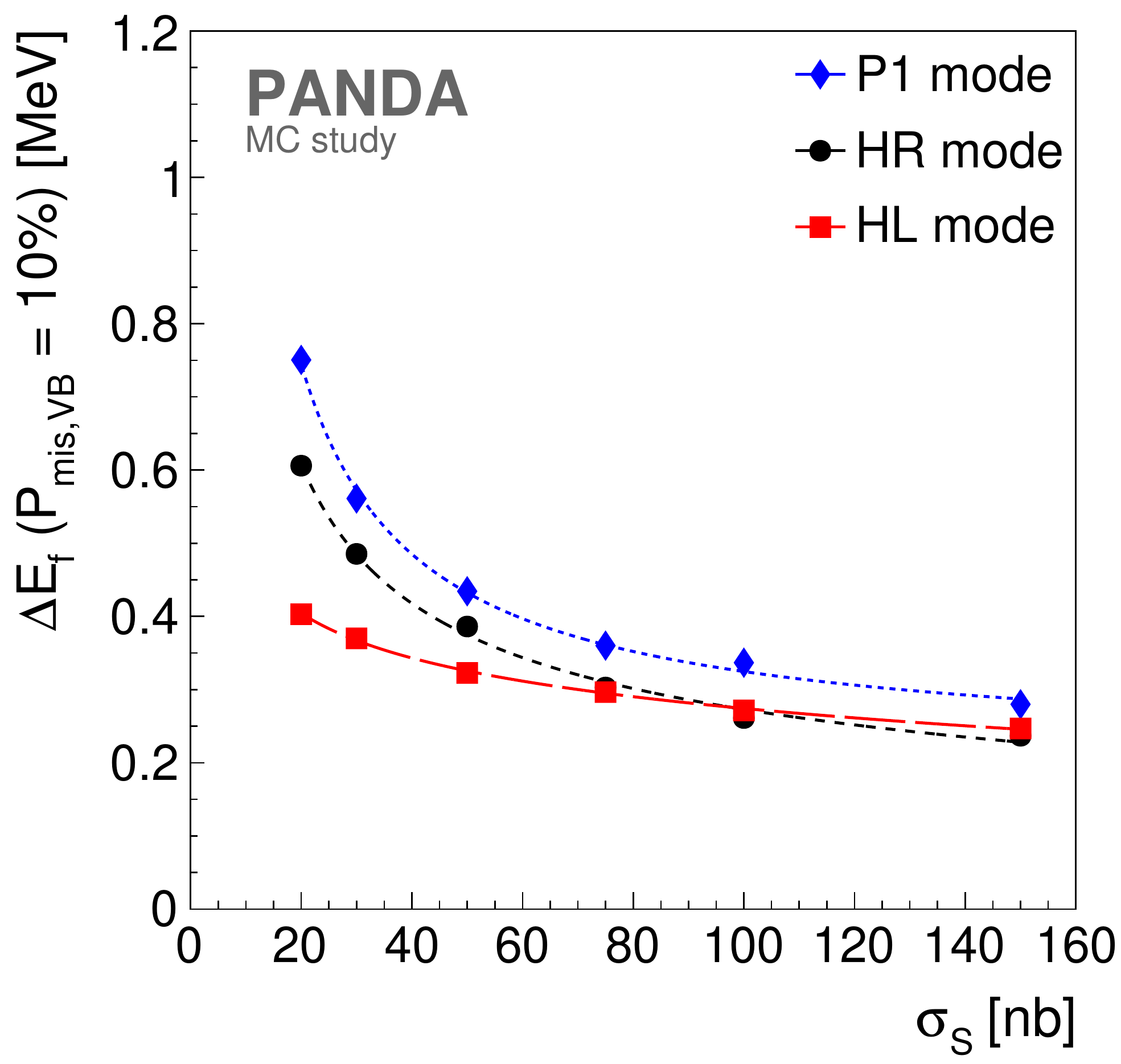} 
  \caption{Sensitivity plot for a misidentification $P_{\rm mis}=10\%$ for the molecular line shape measurement,
    {\it (left)} bound molecular state wrongly identified to be a virtual state ($P_{\rm mis,BV}$) and {\it (right)} virtual state wrongly 
    identified to be a bound molecular state ($P_{\rm mis,VB}$).}
  \label{fig:scanResult_3sigmaBW_AND_10percent_Molecule_trend_bv}
\end{figure*}
The results of the performed sensitivity study for measuring the natural decay width $\Gamma_0$ of a Breit-Wigner like line shape of the $X(3872)$ are comprehensively summarised in Fig.\,\ref{fig:SensitivityResultsBW}, showing the obtained relative precision 
\linebreak
$\Delta\Gamma_{\rm meas}/\Gamma_{\rm meas}$ (Eq.\,\ref{eq:BW_Sens}) {\it vs.} the input width $\Gamma_0$ for the different HESR operation modes (P1, HR, HL) and signal production cross-sections $\sigma_{S}$ under study. 

As it might be expected, the larger the input parameter $\Gamma_0$ and the input cross-section $\sigma_{S}$, the higher is the precision of the measurement. For {\it e.g.} the anticipated HL mode and $\sigma_{S}=50$\,nb, the achievable, interpolated relative uncertainty is about $\leq 10\%$ for $\Gamma_0 \geq 200$\,keV. Here, the best performance is obtained as compared to the other HESR operation modes P1 and HR for about $\Gamma_0 \geq 250$\,keV. For smaller widths ($\Gamma_0 < 250$\,keV), the HR mode delivers superior precision of the width measurements, whereas P1 shows clearly an inferior performance over the full range, as expected. 

In general, the most precise absolute width measurements are obtained in the HR accelerator running mode for smaller input widths of roughly less than $\Gamma_0 \approx 200$\,keV, only moderately depending on the given signal cross-section assumptions $\sigma_{S}$ studied. Otherwise, for input widths larger than $\Gamma_0 \approx 200$\,keV, the measurements based on the HL mode perform best, independently of the given $\sigma_{S}$. Even though based on the P1 mode (with reduced luminosity and beam momentum resolution initially available) results of lowest precision are obtained for absolute width measurements, still a sub-MeV resolution is achieved over the full range of the covered ($\Gamma_0$, $\sigma_{S}$) parameter space. 

A compressed representation of the resultant sensitivities for the Breit-Wigner Case is shown in Fig.\,\ref{fig:scanResult_3sigma_BW_trend}. Extracted from Fig.\,\ref{fig:SensitivityResultsBW}, the minimum $\Gamma_0$, for which a $3\sigma$ sensitivity (that is a $\Delta\Gamma_{\rm meas}/\Gamma_{\rm meas}=33\%$ precision) is achieved in the absolute decay width measurement is plotted {\it vs.} the input $\sigma_{S}$. Trendlines for inter- and extrapolation have again been added using some empirical analytical function.

The sensitivity results for the distinction between 
\linebreak 
bound {\it vs.} virtual nature of the $X(3872)$ via the Flatt\'{e} parameter $E_{\rm f}$ are summarised in Fig.\,\ref{fig:SensitivityResultsEf}, where the misidentification probabilities $P_{\rm mis}$ (Eq.\,\ref{eq:Mol_Sens}) {\it vs.} the assumed input $E_{\rm f,0}$ parameter are presented for the different HESR operation mode scenarios and signal production cross-sec{-}tions $\sigma_{S}$. The higher the difference between the assumed input $E_{\rm f,0}$ value and the threshold value $E_{\rm f,th}$, and also, the higher the assumed $\sigma_{S}$, the better the achievable performance is. 

The $P_{\rm mis}$ expected for {\it e.g.} the anticipated HL mode and $\sigma_{S}=50$\,nb, turns out to be about $\leq 10\%$ for {\it e.g.} $\Delta E_{\rm f} := |E_{\rm f,0} - E_{\rm f,th}|\; \geq |({\rm -}8.91)\;-\;({\rm -}8.56)|\,{\rm MeV} \approx 350\,$keV. Here, for $\sigma_{S}=50$\,nb, the best performance is achieved for the HL mode. This holds for the whole range of input values $E_{\rm f,0}$. 

The HL accelerator running mode offers basically the best performance for the Molecule Case studies. Only for larger signal cross-sections $\sigma_{S} \geq 75$\,nb, the misidentifica{-}tion probability of a virtual state being wrongly assigned to be a bound state $P_{\rm mis,VB}$ ({\it i.e.} data points for $E_{\rm f,0} > E_{\rm f,th}$), the HR mode appears to perform slightly better for small $\Delta E_{\rm f}$ values ({\it e.g.} $\Delta E_{\rm f}\approx 250\,$keV for $\sigma_{S} = 75$\,nb).

Even though based on the P1 mode, the misidentifica{-}tion probabilities achieved are the largest, still a correct identification probability $P_{\rm cor}$ $:=1-P_{\rm mis}$ of better than about $90\%$ for the nature type distinction is feasible for the investigated signal cross-sections $\sigma_{S} \leq 50$\,nb, as long as the $\Delta E_{\rm f}$ values are larger than about $775$\,keV (for input bound state, means $E_{\rm f,0} < E_{\rm f,th}$) and about $455$\,keV (for input virtual state, means $E_{\rm f,0} > E_{\rm f,th}$), respectively.

Also for the Molecule Case, the $P_{\rm mis}$ sensitivity results are more compact represented (Fig.\,\ref{fig:scanResult_3sigmaBW_AND_10percent_Molecule_trend_bv}) in terms of $\Delta E_{\rm f}$ {\it vs.} $\sigma_{S}$ extracted from Fig.\,\ref{fig:SensitivityResultsEf} at $P_{\rm mis}=10\%$. These are shown here separated for both cases, the misidentification of a bound state being wrongly assigned to be a virtual state $P_{\rm mis,VB}$ (Fig.\,\ref{fig:SensitivityResultsEf}, {\it left}) and the one of a virtual state being wrongly assigned to be a bound state $P_{\rm mis,VB}$ (Fig.\,\ref{fig:SensitivityResultsEf}, {\it right}), together with some empirical trendlines for inter- and extrapolation.

For completeness, the full set of sensitivity numbers is also summarised in Tabs.\,\ref{tab:sensi_BW_P1_HR_HL} and \ref{tab:sensi_Pmis_Molecule_P1_HR_HL} in the Appendix.

\section{Summary}
\label{sec:Summary}
A comprehensive feasibility study for resonance energy scans of narrow states at PANDA has been carried out. Using the example of the famous, presumably exotic $X(3872)$ state, experiments for absolute decay width and line shape measurements have been realistically simulated, analysed and the projected performances are quantified.

For both cases, the Breit-Wigner and the Molecule Case, and three anticipated HESR accelerator running modes (P1, HL, HR), the resonance energy scan experiments have been studied each for 40 energy scan points and an assumed data-taking time period of two days per point. Based on the nominal HESR performances, including a realistic estimate of involved uncertainties, the achievable performances have been provided using for the physics input either experimental measurements where available or realistic assumptions according to the example state $X(3872)$ and presently running experiments.

The outcome of the sensitivity study for the Breit-Wigner Case for an input signal cross-section of $\sigma_{S} = 50$\,nb, in line with the experimental upper limit on 
\linebreak
$X(3872) \rightarrow p\bar{p}$ provided by the LHCb experiment~\cite{LHCbXppUpdate}, can exemplarily be summarised as follows. A $3\sigma$ precision, $\Delta\Gamma_{\rm meas}/\Gamma_{\rm meas}$ better than 33\%, is achieved for an assumed natural decay width larger than about $\Gamma_0 = 40$\,keV (HR), $\Gamma_0 = 80$\,keV (HL) and $\Gamma_0 = 110$\,keV (P1), respectively.
We find the HR mode superior for very narrow widths smaller than about $\Gamma_0 = 180$\,keV over the investigated $\sigma_{S}$ range of [20 - 150]\,nb. A proof of principle for an experimental distinction between the bound {\it vs.} virtual nature with the PANDA experiment based on measurements of line shapes characterised by the dynamical Flatt\'{e} parameter $E_{\rm f}$ has been provided.
 
The achievable sensitivities in terms of misidentifica{-}tion probability $P_{\rm mis}$ for such molecular line shape measurements have been quantified. They can be summarised, again for one (out of the six) assumed $\sigma_{S} = 50$\,nb, as follows. A 90\% probability to correctly identify the nature (bound or virtual) of the state for $\Delta E_{\rm f}$ larger than 300\,keV (HL), 400\,keV (HR) and 700\,keV (P1) can be expected, respectively. We find the HL mode superior mostly over the full investigated $\sigma_{S}$ range of [20 - 150]\,nb.

\section{Acknowledgements}
\label{sec:Acknow}
We acknowledge financial support from the Science and Technology Facilities Council (STFC), 
British funding a{-}gency, Great Britain; the Bhabha Atomic Research Centre (BARC) and the 
Indian Institute of Technology Bombay, India; the Bundesministerium f\"ur Bildung und 
Forschung (BMBF), Germany; the Carl-Zeiss-Stiftung 21-0563-2.8/{-}122/1 and 21-0563-2.8/131/1, 
Mainz, Germany; the Center for Advanced Radiation Technology (KVI-CART), 
\linebreak
Groningen, Netherlands; 
the CNRS/IN2P3 and the Universit\'{e} Paris-Sud, France; 
the Czech Ministry (MEYS) grants LM2015049, CZ.02.1.01/0.0/0.0/16 and 013/000
\linebreak
1677, the Deutsche Forschungsgemeinschaft (DFG), 
Germany; the Deutscher Akademischer Austauschdienst 
\linebreak 
(DAAD), Germany; the Forschungszentrum 
J\"ulich, Germany; the FP7 HP3 GA283286, European Commission funding; the Gesellschaft f\"ur 
Schwerionenforschung GmbH (GSI), 
Darmstadt, Germany; the Helmholtz-Gemeinschaft 
\linebreak
Deutscher 
Forschungszentren (HGF), Germany; the INTAS, European Commission funding; the Institute of High 
Energy Physics (IHEP) and the Chinese Academy of Sciences, Beijing, China; the Istituto 
Nazionale di Fisica Nucleare (INFN), Italy; the Ministerio de Educacion y Ciencia (MEC) under 
grant FPA2006-12120-C03-02; the Polish Ministry of Science and Higher Education (MNiSW) grant 
No. 2593/7, PR UE/2012/2, and the National Science Centre (NCN) DEC-2013/09/N/ST2/02180, Poland; 
the State Atomic Energy Corporation Rosatom, National Research Center Kurchatov Institute, Russia; 
the Schweizerischer Nationalfonds zur F\"orderung der Wissenschaft{-}lichen Forschung (SNF), Swiss; 
the Stefan Meyer Institut f\"ur Subatomare Physik and the \"Osterreichische Akademie der Wissenschaften, 
Wien, Austria; the Swedish Research Council and the Knut and Alice Wallenberg Foundation, Sweden.
\clearpage
\section{Appendix}
\label{sec:Appdx} 
%
%
\begin{table*}[h!]
\caption{{\it Top:} Sensitivities $\Delta(\Gamma / \Gamma_{\rm meas})$ [\%] for the Breit-Wigner Case (P1/HR/HL). 
{\it Bottom:} Sensitivity uncertainties due to generic hadronic and non-resonant background contributions, based on upscaling [\%].}
\vspace{1ex}
\label{tab:sensi_BW_P1_HR_HL}
\centering
\begin{small}
\begin{tabular}{c|cccccc}
$\Gamma_0$ [keV] &   20nb &   30nb &   50nb &   75nb &  100nb &  150nb\\ \hline \\
   50  &  92.0\,/\,59.2\,/\,77.7  &  84.9\,/\,41.7\,/\,71.9  &  68.4\,/\,28.3\,/\,60.6  &  57.5\,/\,19.6\,/\,55.1  &  51.5\,/\,15.9\,/\,46.4  &  39.3\,/\,12.4\,/\,34.4 \\ 
   70  &  84.8\,/\,41.9\,/\,69.2  &  74.3\,/\,29.1\,/\,61.4  &  56.2\,/\,18.8\,/\,46.7  &  43.2\,/\,14.5\,/\,35.0  &  34.5\,/\,11.9\,/\,28.0  &  25.8\,/\,9.1\,/\,21.4  \\ 
  100  &  68.6\,/\,30.7\,/\,55.6  &  52.6\,/\,20.8\,/\,43.5  &  39.4\,/\,14.9\,/\,28.3  &  27.5\,/\,11.2\,/\,20.7  &  22.8\,/\,9.7\,/\,16.8   &  15.7\,/\,7.0\,/\,11.9  \\ 
  130  &  59.7\,/\,27.7\,/\,43.6  &  44.8\,/\,19.1\,/\,28.6  &  29.2\,/\,12.9\,/\,19.4  &  21.0\,/\,9.9\,/\,14.5   &  17.3\,/\,8.1\,/\,11.4   &  13.0\,/\,6.1\,/\,8.5   \\ 
  180  &  47.0\,/\,22.7\,/\,27.4  &  31.6\,/\,16.9\,/\,19.2  &  20.9\,/\,11.6\,/\,12.9  &  15.5\,/\,8.5\,/\,9.2    &  12.7\,/\,7.0\,/\,7.5    &  9.6\,/\,5.7\,/\,5.7    \\ 
  250  &  34.0\,/\,21.1\,/\,18.5  &  24.7\,/\,15.0\,/\,13.6  &  16.4\,/\,10.8\,/\,8.9   &  12.6\,/\,8.3\,/\,6.3    &  10.2\,/\,6.6\,/\,5.2    &  7.5\,/\,5.0\,/\,3.8    \\ 
  500  &  25.1\,/\,18.9\,/\,10.0  &  17.6\,/\,13.4\,/\,6.6   &  11.9\,/\,9.5\,/\,4.6    &  9.5\,/\,7.3\,/\,3.7     &  7.6\,/\,6.3\,/\,2.9     &  5.9\,/\,4.8\,/\,2.3    \\ 
\\
\hline
\\
   50  &  94.0\,/\,77.1\,/\,80.5 &  91.3\,/\,57.5\,/\,79.8 &  76.3\,/\,38.7\,/\,70.7 &  67.2\,/\,27.0\,/\,64.7 &  63.8\,/\,21.5\,/\,58.5  &  48.7\,/\,15.4\,/\,45.1 \\ 
   70  &  86.5\,/\,64.9\,/\,79.3 &  82.6\,/\,42.8\,/\,71.8 &  65.4\,/\,25.9\,/\,58.7 &  52.1\,/\,18.3\,/\,50.3 &  45.3\,/\,15.4\,/\,38.2  &  34.7\,/\,11.2\,/\,29.1 \\ 
  100  &  80.5\,/\,46.0\,/\,67.0 &  67.8\,/\,31.7\,/\,57.1 &  51.6\,/\,20.7\,/\,41.7 &  37.7\,/\,14.3\,/\,30.5 &  28.0\,/\,11.3\,/\,23.6  &  21.9\,/\, 8.5\,/\,16.1 \\ 
  130  &  77.4\,/\,43.4\,/\,57.2 &  61.5\,/\,26.6\,/\,45.2 &  39.0\,/\,17.5\,/\,28.4 &  28.5\,/\,12.3\,/\,19.9 &  21.6\,/\,9.9\,/\,15.7   &  16.3\,/\,6.9\,/\,11.2 \\ 
  180  &  62.6\,/\,33.9\,/\,42.2 &  44.6\,/\,22.4\,/\,29.3 &  28.3\,/\,15.1\,/\,18.9 &  20.1\,/\,10.5\,/\,12.4 &  16.0\,/\,8.8\,/\,9.9    &  11.7\,/\,6.4\,/\,7.1 \\ 
  250  &  51.3\,/\,29.7\,/\,28.7 &  34.5\,/\,20.6\,/\,19.5 &  21.6\,/\,13.4\,/\,12.2 &  15.6\,/\,9.8\,/\,8.5   &  12.2\,/\,8.0\,/\,6.8    &   9.1\,/\,5.9\,/\,4.8 \\ 
  500  &  34.2\,/\,28.1\,/\,14.2 &  23.0\,/\,19.9\,/\,10.1 &  15.5\,/\,11.9\,/\,6.3 &   11.1\,/\,9.0\,/\,4.6   &   9.3\,/\,7.1\,/\,3.8    &   6.5\,/\,5.5\,/\,2.8 \\ 
\end{tabular}
\end{small}
\vspace{0.3cm}
\end{table*}
~\\\\\\
\vspace{0.6cm}
\begin{table*}[h!]
\caption{{\it Top:} Sensitivities in terms of misidentification probabilities $P_{\rm mis}$ [\%] for the Molecule Case (P1/HR/HL).
{\it Bottom:} Sensitivity upper limits due to generic hadronic and non-resonant background contributions, based on upscaling [\%].}
\vspace{1ex}
\label{tab:sensi_Pmis_Molecule_P1_HR_HL}
\centering
\begin{small}
\begin{tabular}{c|cccccc}
$E_{f0}$ [MeV] &   20nb &   30nb &   50nb &   75nb &  100nb &  150nb\\ \hline \\
  -10.0  &   4.1\,/\,0.9\,/\,0.3      &   2.5\,/\,0.5\,/\,0.1       &   0.8\,/\,0.0\,/\,0.1     &   0.1\,/\,0.0\,/\,0.0     &   0.2\,/\,0.0\,/\,0.0     &   0.0\,/\,0.0\,/\,0.0    \\ 
   -9.5  &  10.4\,/\,4.4\,/\,2.9      &   8.5\,/\,2.7\,/\,1.2       &   3.1\,/\,0.3\,/\,0.4     &   0.6\,/\,0.0\,/\,0.0     &   0.2\,/\,0.0\,/\,0.0     &   0.0\,/\,0.0\,/\,0.0    \\ 
   -9.0  &  29.1\,/\,19.0\,/\,18.0    &  26.5\,/\,12.8\,/\,10.3     &  19.1\,/\,7.1\,/\,4.7     &  11.9\,/\,3.3\,/\,2.4     &   7.4\,/\,1.9\,/\,0.9     &   3.0\,/\,0.6\,/\,0.1    \\ 
   -8.8  &  44.3\,/\,36.7\,/\,31.5    &  37.1\,/\,30.3\,/\,23.1     &  34.6\,/\,23.5\,/\,15.4   &  28.1\,/\,14.2\,/\,9.6    &  20.4\,/\,10.0\,/\,6.9    &  14.8\,/\,6.4\,/\,3.4    \\ 
   -8.3  &  27.9\,/\,28.1\,/\,20.6    &  25.2\,/\,22.8\,/\,19.1     &  19.4\,/\,19.0\,/\,14.5   &  17.4\,/\,12.7\,/\,12.9   &  15.7\,/\,9.7\,/\,10.7    &  11.2\,/\,8.0\,/\,8.7    \\ 
   -8.0  &  17.7\,/\,11.2\,/\,3.4     &  11.9\,/\,7.2\,/\,2.0       &   5.4\,/\,3.1\,/\,1.4     &   2.2\,/\,1.4\,/\,0.5     &   1.6\,/\,1.7\,/\,0.1     &   0.6\,/\,0.5\,/\,0.0    \\ 
   -7.5  &   4.6\,/\,1.9\,/\,0.0      &   0.9\,/\,0.5\,/\,0.0       &   0.0\,/\,0.0\,/\,0.0     &   0.0\,/\,0.0\,/\,0.0     &   0.0\,/\,0.0\,/\,0.0     &   0.0\,/\,0.0\,/\,0.0    \\ 
   -7.0  &   0.8\,/\,0.5\,/\,0.0      &   0.0\,/\,0.2\,/\,0.0       &   0.0\,/\,0.0\,/\,0.0     &   0.0\,/\,0.0\,/\,0.0     &   0.0\,/\,0.0\,/\,0.0     &   0.0\,/\,0.0\,/\,0.0    \\ 
\\
\hline 
\\
  -10.0  &   8.4\,/\,2.2\,/\,3.3      &   4.8\,/\,1.3\,/\,0.6       &   1.7\,/\,0.1\,/\,0.2     &   0.3\,/\,0.0\,/\,0.0     &   0.4\,/\,0.0\,/\,0.0     &   0.1\,/\,0.0\,/\,0.0    \\ 
   -9.5  &  20.3\,/\,7.6\,/\,8.1      &  11.5\,/\,5.1\,/\,3.7       &   6.9\,/\,1.6\,/\,0.9     &   2.8\,/\,0.0\,/\,0.4     &   0.9\,/\,0.3\,/\,0.2     &   0.1\,/\,0.1\,/\,0.1    \\ 
   -9.0  &  34.3\,/\,26.8\,/\,25.7    &  31.1\,/\,19.1\,/\,19.2     &  23.5\,/\,12.7\,/\,10.0   &  18.8\,/\,7.9\,/\,4.3     &  12.9\,/\,3.8\,/\,2.3     &   7.0\,/\,1.9\,/\,1.4    \\ 
   -8.8  &  46.8\,/\,38.4\,/\,43.5    &  41.1\,/\,36.8\,/\,30.4     &  37.5\,/\,27.2\,/\,24.5   &  34.4\,/\,23.0\,/\,14.7   &  29.2\,/\,15.8\,/\,9.0    &  19.7\,/\,11.5\,/\,6.9   \\ 
   -8.3  &  31.5\,/\,29.2\,/\,27.3    &  28.0\,/\,26.8\,/\,22.1     &  24.0\,/\,22.5\,/\,18.3   &  20.4\,/\,19.8\,/\,15.4   &  19.1\,/\,14.5\,/\,12.9   &  14.7\,/\,9.4\,/\,13.2   \\ 
   -8.0  &  19.3\,/\,17.2\,/\,9.2     &  16.8\,/\,12.6\,/\,4.4      &  11.2\,/\,7.2\,/\,1.6     &   6.2\,/\,2.6\,/\,0.8     &   3.4\,/\,3.7\,/\,1.0     &   1.2\,/\,0.8\,/\,0.1    \\ 
   -7.5  &   9.7\,/\,7.3\,/\,0.5      &   3.4\,/\,2.2\,/\,0.0       &   0.5\,/\,0.1\,/\,0.0     &   0.1\,/\,0.1\,/\,0.0     &   0.0\,/\,0.0\,/\,0.0     &   0.0\,/\,0.0\,/\,0.0    \\ 
   -7.0  &   3.1\,/\,2.0\,/\,0.0      &   0.2\,/\,0.6\,/\,0.0       &   0.0\,/\,0.0\,/\,0.0     &   0.0\,/\,0.0\,/\,0.0     &   0.0\,/\,0.0\,/\,0.0     &   0.0\,/\,0.0\,/\,0.0    \\ 
\end{tabular}
\end{small}
\end{table*}

\clearpage

%

\end{document}